\documentstyle[12pt]{article}

\title{Darboux transforms on band matrices, weights and associated polynomials}

\author{Mark Adler\thanks{Appears in: Intern. Math.
Research Notices, 2001.
 \newline Department of Mathematics, Brandeis University,
Waltham, Mass 02454, USA. E-mail:
adler@math.brandeis.edu. The support of a National
Science Foundation grant \# DMS-98-4-50790 is
gratefully acknowledged.}~~~~~~ Pierre van
Moerbeke\thanks{Department of Mathematics,
Universit\'e de Louvain, 1348 Louvain-la-Neuve,
Belgium and Brandeis University, Waltham, Mass 02454,
USA. E-mail: vanmoerbeke@geom.ucl.ac.be and
@brandeis.edu. The  support of a National Science
Foundation grant \# DMS-98-4-50790, a Nato, a FNRS and
a Francqui Foundation grant is gratefully
acknowledged.}}

\date{November 17, 2000}

\newcommand{\MAT}[1]{\left(\begin{array}{*#1c}}
\newcommand{\mat}{\end{array}\right)}
\newcommand{\qed}
{%
\mbox{}%
\nolinebreak%
\hfill%
\rule{2mm}{2mm}%
\medbreak%
\par%
}

\newcommand{\sumbis}[2]%
{%

\begin{array}[t]{c}
\sum\\
{\scriptstyle #1}\\
{\scriptstyle #2}
\end{array}

}

\newcommand{\LR}{{\cal L}}

\newcommand{\BC}{{\Bbb C}}

\newcommand{\BX}{{\Bbb X}}

\newcommand{\BZ}{{\Bbb Z}}

\newcommand{\iy}{\infty}
\newcommand{\pl}{\partial}
\newcommand{\al}{\alpha}
\newcommand{\proof}{\underline{\sl Proof}: }
\newcommand{\remark}{\underline{\sl Remark}: }
\newcommand{\example}{\underline{\sl Example}: }
\newcommand{\om}{\omega}

\newcommand{\SR}{{\cal S}}

\newcommand{\la}{\langle}
\newcommand{\ra}{\rangle}

\newcommand{\dt}{\delta}

\newcommand{\BR}{{\Bbb R}}
\newcommand{\lb}{\lambda}
\newcommand{\Lb}{\Lambda}

\def\span{\mathop{\rm span}}
\def\diag{\mathop{\rm diag}}

\def\be{\begin{equation}}
\def\ee{\end{equation}}
\def\bea{\begin{eqnarray}}
\def\eea{\end{eqnarray}}
\def\bean{\begin{eqnarray*}}
\def\eean{\end{eqnarray*}}

\ifx\undefined\Bbb
        \let\Bbb\bf
\fi

\catcode`\@=11
\def\ps@X{\let\@mkboth\@gobbletwo
        \def\@oddhead{\tt Adler-van Moerbeke:%
        Band matrix\hfil\ \ Nov 17, 2000%\today
        \ \ \hfil\S\thesection,
p.\thepage
        }
        \def\@oddfoot{\rm\hfil\thepage\hfil}
        \let\@evenhead\@oddhead
        \let\@evenfoot\@oddfoot}
\catcode`@=12
\catcode`\@=11
\let\c@equation=\relax
\newcounter{equation}[subsection]

\catcode`\@=12

\newtheorem{definition}{Definition}[%sub
section]

\newtheorem{theorem}[definition]{Theorem}

\newtheorem{lemma}[definition]{Lemma}
\newtheorem{corollary}[definition]{Corollary}
\newtheorem{proposition}[definition]{Proposition}

\catcode`\@=11
\let\c@equation=\relax
\newcounter{equation}[%sub
section]

\catcode`\@=12

\begin{document}
\maketitle

\begin{abstract} Classically, it is well known
 that a single weight on
a real interval leads to orthogonal polynomials. In
"Generalized orthogonal polynomials, discrete KP
 and Riemann-Hilbert problems", Comm. Math. Phys.
  207, pp. 589-620 (1999), we have shown that $m$-periodic
sequences of weights lead to "moments", polynomials
defined by determinants of matrices
 involving these moments and
$2m+1$-step relations between them, thus leading to
$2m+1$-band matrices $L$. Given a Darboux transformations
 on $L$, which effect does it have on the
$m$-periodic sequence of weights and
on the associated polynomials ?  These questions will
receive a precise answer in this paper. The methods are based on
introducing time parameters in the weights,
making the band matrix $L$ evolve according to the
so-called
discrete KP hierarchy. Darboux transformations on that
$L$ translate into vertex operators acting on the
$\tau$-function.

\end{abstract}

\setcounter{equation}{0}

%\vspace{0.5cm}
\tableofcontents

\vspace{1cm}

{\bf Classical situation: a weight and tridiagonal
matrices}. A single weight $\rho(z),~z \in \BR$,
 naturally leads to a moment matrix,
  $$
m_n=\left(\mu_{ij}\right)_{0\leq i,j \leq n-1}
=
 \left(  \la z^i,z^j \rho(z)\ra\right)_{0\leq i,j
\leq n-1}=
 \left(  \la z^i,  \rho_j(z)\ra\right)_{0\leq i,j
\leq n-1},
  $$
  where $\la
f,g \ra = \int _{\BR} fg dz$ and where
$\rho_j(z):=z^j\rho(z)$. In turn, the moments lead to
a sequence of monic orthogonal polynomials
\begin{eqnarray}
 p_n(z)&=&\frac{1}{\det m_n}\det\left(
 \begin{tabular}{lll|l}
$\mu_{00}$&...&$\mu_{0,n-1}$&1\\ $\vdots$& &$\vdots$&$\vdots$\\
$\mu_{n-1,0}$&...&$\mu_{n-1,n-1}$&$z^{n-1}$\nonumber\\
\hline
$\mu_{n0}$&...&$\mu_{n,n-1}$&$z^n$
\end{tabular}\right),\nonumber\\\nonumber
\end{eqnarray}
thus satisfying $$\int_{\BR}
p_k(z)p_{\ell}(z)\rho(z)dz=\delta_{k\ell}h_k. $$ Then,
as is classically well known, the vector
$p(z)=(p_0(z),p_1(z), p_2(z),...)$ of polynomials
leads to tridiagonal matrices $L$, defined by
$zp(z)=Lp(z)$.

\bigskip

{\bf Periodic sequences of weights and $2m+1$-band
matrices}. Instead of the classical situation, where
$\rho_j(z)=z^j\rho(z)$, we consider an ``{\em
$m$-periodic}" sequence of weights
$\rho(z):=(\rho_j(z))_{j\geq 0}$ on $\BR$; i.e.,
satisfying
\be
z^m \rho_j(z)=\rho_{j+m}(z);
 \ee
 in other words,
 \be
  \rho =
 \left( \rho_0, \rho_1,..., \rho_{m-1}, z^m
  \rho_0,..., z^m  \rho_{m-1},z^{2m} \rho_0
 ,..., z^{2m}  \rho_{m-1},
 ...
  \right).
  \ee
 This leads naturally to a $2m+1$-band matrix!  Indeed, to this sequence and the inner-product $\la f,g \ra = \int _{\BR} fg dz$,
we associate, by analogy, the semi-infinite ``{\em
moment matrix}" $m_{\iy}(\rho)$, where
\be
m_{n}(\rho):=\left( \mu_{ij}(\rho)\right)_{0\leq i,j
\leq  n-1}
 :=\left(
\left\la z^i,\rho_j(z)\right\ra \right)_{0\leq i,j
\leq n-1},\ee
 the determinant
   $$
  D_n (\rho):=\det m_n (\rho),
   $$
  and the infinite sequence of monic
polynomials, where $\mu_{ij}=\mu_{ij}(\rho)$,
\begin{eqnarray}
 p_n(z)&=&\frac{1}{ D_n (\rho)}\det\left(
 \begin{tabular}{lll|l}
$\mu_{00}$&...&$\mu_{0,n-1}$&1\\ $\vdots$& &$\vdots$&$\vdots$\\
$\mu_{n-1,0}$&...&$\mu_{n-1,n-1}$&$z^{n-1}$\nonumber\\
\hline
$\mu_{n0}$&...&$\mu_{n,n-1}$&$z^n$
\end{tabular}\right)\nonumber\\
&=&\frac{1}{ D_n (\rho)}
 \det(z \mu_{ij}-\mu_{i+1,j})_{0\leq i,j \leq n-1}
 .
\end{eqnarray}
 The second formula for $p_n(z)$ will be discussed in
Lemma 2.2. Throughout the paper, the $D_n(\rho)$'s are
assumed to be non-zero. Then the sequence $p_n(z)$
gives rise to a semi-infinite matrix $L$, defined by
\be
z^m p(z)=L p(z), \ee where $L$ is a  {\em $2m+1$-band
matrix}\footnote{zero everywhere, except for $m$
consecutive subdiagonals on either side of the main
diagonal.}; this was established by us in \cite{AvM6}
and a sketch of the proof will be given in
 Proposition 2.3.
Moreover, Gr\"unbaum and Haine \cite{GH} had produced
a sequence of~ "$5$-step polynomials", satisfying a
fourth order differential equation and related to the
classical Krall orthonormal polynomials. As we shall
see, these polynomials are very special cases of our
theory. We conjecture that all sequences of
polynomials satisfying $2m+1$-step relations of the
precise form (0.5) are given by {\em generalized
periodic sequences of weights}, a slight
generalization of (0.1), and limiting cases thereof.
See definition 3.2.

\medbreak

In Theorems 0.1 and 0.2, we shall be conjugating with
the following matrices:
 $$
\beta \Lb^0+\Lb
=
 \left(
\begin{array}{ccccc}
 \beta_0&1&0&0&  \\
 0&\beta_1&1&0&\\
 0&0&\beta_2&1&\\
 0&0&0&\beta_3&\\
 & & & &\ddots\\
\end{array}  \right)
~~\mbox{ and }~
\Lb^{\top}\beta +I=\left(
\begin{array}{ccccc}
 1&0&0&0&  \\
 \beta_0&1&0&0&\\
 0&\beta_1&1&0&\\
 0&0&\beta_2&1&\\
 & & & &\ddots\\
\end{array}  \right) ,
$$ where $\Lb$ is the semi-infinite
 shift matrix $\Lambda
:=(\dt_{i,j-1})_{i,j\geq 0}$, i.e., $(\Lambda
v)_n=v_{n+1}$. Note, in the semi-infinite case, $\Lb
\Lb^{\top}=I\neq \Lb^{\top}\Lb$.

%\newpage

\begin{theorem}{\bf (LU-Darboux transforms)} The Lower-Upper
Darboux transform
 \be
  L-\lb^m I\longmapsto \tilde L-\lb^m
I:=(\beta\Lb^0+\Lb)(L-\lb^m I)(\beta\Lb^0+\Lb)^{-1}
 \ee
maps $L$ into a new $2m+1$-band matrix $\tilde L$,
provided
$$
\beta_n=-\frac{\Phi_{n+1}(\lb)}{\Phi_n(\lb)}
 ~~\mbox{with arbitrary}~~
 \Phi(\lb)=(\Phi^{}_n(\lb))_{n\geq 0}
 %(\Phi_0(\lb),\Phi_1(\lb),\ldots)
 \in
  (L-\lb^m I)^{-1}
  (0,0,...).
$$ The null-space $ (L-\lb^m I)^{-1}
  (0,0,...)$ is $m$-dimensional with basis vectors
% $\Phi(\lb)$ is an arbitrary linear combination of
%the following basis vectors, for $1\leq k\leq n$,
 given by
  $$
\Phi^{(k)}(\lb) = \left( \frac{
 D_n\Bigl(\tilde\rho^{(k)}\Bigr)}{D_n (\rho)}
\right)_{n\geq 0},~~~\mbox{for $1\leq k\leq m$},
  $$
where \be
 \tilde \rho^{(k)} (z) := (\om^k \lb - z)
  \rho (z)= \left((\om^k\lb-z)\rho_0(z),~
 (\om^k\lb-z)\rho_1(z),\ldots\right)
 .\ee
   The LU-Darboux transformation $ L-\lb^m I\longmapsto \tilde
L-\lb^m I$
 associated with each
$$
 \beta_n = -\frac{\Phi^{(k)}_{n+1}(\lb)}{\Phi^{(k)}_n(\lb)} ,
 ~~\mbox{for fixed}~~1\leq k \leq m ,
 $$
 induces a map on $m$-periodic weights
 \be
 \rho (z)%=(\rho_0,\rho_1,\ldots)
  \longmapsto \tilde \rho^{(k)} (z) ,
  \ee
  with $\tilde \rho^{(k)}$ leading to the $2m+1$-band matrix
$\tilde L$.

\end{theorem}

\remark Section 5 (Theorem 5.1) contains the proof of
a
 more general statement, involving linear combinations of
$\Phi^{(k)}(\lb) $ .

%\newpage

\begin{theorem}{\bf (UL-Darboux transforms)}. The Upper-Lower
Darboux transform
 \be
 L - \lb^m I \longmapsto \tilde L - \lb^m I := (\Lb^{\top} \beta + I)
 (L - \lb^m I) (\Lb^{\top}{\beta} + I)^{-1} ,
 \ee
 maps $L$ into a new $2m+1$-band matrix $\tilde L$,
provided\footnote{$e_i:=(0,\ldots,{\underbrace{1}_i
 },0,\ldots)\in \BR^{\iy}$.}% (for the precise meaning, see Theorems 4.1
% and 4.2)
 $$
 \beta_n = - \frac{\Phi_{n+1} (\lb)}{\Phi_n (\lb)}~~\mbox{with}~~
 \Phi(\lb)=
 (\Phi^{}_n(\lb))_{n\geq 0}\in (L-\lb^m I)^{-1} \span (e_1,e_2,\ldots,e_m)
  %({\underbrace{ *,*,...,*}_{m}},0,0,...)
 . $$
The(quasi)-null vectors $\Phi(\lb)$ of $L-\lb^m I$
depends projectively on $2m-1$ free parameters
$a_0,...,a_{m-1},b_0,...,b_{m-1}$ \footnote{The
UL-Darboux transform depends on $m$ additional free
parameters, compared to the LU transform .} and are
given by
 \be
\Phi^{}(\lb) = \left((-1)^{n-1}
\frac{
 D_{n+1}(\tilde\rho )}{D_n (\rho)}
\right)_{n\geq 0},%~~~\mbox{for $1\leq k\leq m$},
  \ee
where
 \be
 \tilde \rho =
 \left(\tilde \rho_0,\tilde \rho_1,...,\tilde \rho_{m-1}, z^m
 \tilde \rho_0,..., z^m \tilde \rho_{m-1},z^{2m}\tilde \rho_0
 ,..., z^{2m} \tilde \rho_{m-1},
 ...
  \right),
  \ee
with\footnote{the delta-function is defined in the
standard way $\int f(z)\delta(\lb-z)dz=f(\lb)$.}
 \bea
  \tilde \rho_0 (z) :&=&
 \sum_{k=0}^{m-1} \left( a_k \dt (z - \om^k \lb) +
  b_k \frac{\rho_k (z)}
 {z^m - \lb^m }\right),~~\mbox{with}~~ b_{m-1}\neq 0,\nonumber\\
 \tilde \rho_k (z):&=& \rho_{k-1} (z ),
~~~~~~ \mbox{ for } ~~ 1\leq k \leq m-1.
 \eea
 The
UL-Darboux transform $L - \lb^m I \longmapsto \tilde L
- \lb^m I$ induces a map on $m$-periodic sequence of
weights $$\rho%=(\rho_0,\rho_1,\ldots)
\longmapsto \tilde \rho,
  $$
with $\tilde \rho^{}$ leading to the $2m+1$-band
matrix $\tilde L$.
\end{theorem}

\vspace{.2cm}

\begin{corollary}
An appropriate choice of $a_k $, and appropriate limits $b_k \mapsto
 \infty$ and $\lb \mapsto 0$ yield the following special Darboux
 transformation on the $m$-periodic weights
 $$\rho=(\rho_0,\rho_1,\ldots) \mapsto \tilde \rho=(\tilde \rho_0 ,\tilde \rho_1 ,\tilde  \rho_2
,\ldots),$$
 with new weights
 \bean
 \tilde \rho_0 (z):&= &
 \sum_{k=0}^{m-1} \left(c_k \left(\frac{d}{dz}\right)^k
 \dt (z) + d_k \frac{\rho_k (z)}{z^m }\right),~\mbox{with}~d_{m-1}\neq 0,\\
  \tilde \rho_{k} (z) :&=& \rho_{k - 1} (z)~~
  ~~~~\mbox{ for }~~  1\leq k \leq m-1.
 \eean
\end{corollary}

 Weights with $\dt$-functions have been studied mainly by Krall and
Scheffer \cite{KS} and Koornwinder \cite{K}, at least for the
standard orthogonal polynomials. For recent expositions on the
subject, see, for instance,
 Andrews and
Askey \cite{AA}. Recently, they have been studied by
Gr\"unbaum-Haine \cite{GH} and
Gr\"unbaum-Haine-Horozov \cite{GHH}.

\bigbreak

\noindent {\bf An integrable flow with initial $m_{\iy}$ }. We have
 introduced the method of inserting the time in the context
 of random matrices \cite{ASV2,ASV3,vM}, where it has turned out
 to be very useful. In order to establish the results above,
 consider, as we did in \cite{AvM3, AvM4}, the initial
 value problem, depending on two sequences of time parameters
  $x=(x_1,x_2,...)$ and $y=(y_1,y_2,...)$:
\be
 \frac{\pl m_{\iy}}{\pl x_{n}}=\Lb^n m_{\iy},~~
 \frac{\pl m_{\iy}}{\pl y_{n}}=
 -m_{\iy} \Lb^{\top n},
 \mbox{ with initial }m_{\iy}(0,0)=(\la
 z^{i},\rho_{j}(z)\ra)_{0\leq i, j<\iy},
\ee where $\Lb$ is the customary (semi-infinite) shift
matrix. As we shall establish in section 2, imposing
the condition \be~\Lb^m m_{\iy}=m_{\iy}\Lb^{\top m}
\ee on moment matrices $m_{\iy}$ leads to $2m+1$-band
matrices. This in turn, suggests the following useful
reduction: given the times $x,y\in \BC^{\iy}$, we
define new {\em times} $\bar x,\bar y,\bar t
\in
\BC^{\iy},$
 \bean
  \bar
x&=&(x_1,...,x_{m-1},0,x_{m+1},...,
x_{2m-1},0,x_{2m+1},...)
\\
\bar y&=&(y_1,...,y_{m-1},0,y_{m+1},...
,y_{2m-1},0,y_{2m+1},...)\\
\bar t&=&(0,...,0,t_m,0,...,0,t_{2m} ,0,...,0,t_{3m},0,...),
\eean
with
\be
t_{km}:=x_{km}-y_{km}~~~\mbox{ for }~~k=1,2,...  ~.
\ee The point is that, letting $m_{\iy}$ evolve
according to the variables $\bar x,\bar y, \bar t$,
will conserve the $2m+1$-band form of $L$. The
solution to the initial value problem (0.13) is given
by the same moment matrix $m_{\iy}$, as in (0.13),
\be
m_{\iy}\left(\rho(z;\bar x,\bar y, \bar t)\right)
=\left(\left\la z^i, \rho_{j} (z; \bar x, \bar
 y, \bar t)\right\ra\right)_{0\leq i, j<\iy},
\ee but for weights, now depending on times $\bar
x,\bar y, \bar t$, defined as \footnote{The ${\bf
s}_{\ell}$'s denote the elementary Schur polynomials
$e^{\sum_1^{\iy}t_i z^i}=\sum_0^{\iy}{\bf
s}_n(t)z^n$.}
\be
\rho_{j}(z;\bar x,\bar y,\bar t)=
 e^{\sum_1^{\iy} \bar x_{r}z^r}
 e^{\sum_{1}^{\iy} \bar t_{{\ell m}}z^{\ell m}}
\sum_{\ell=0}^{\iy} {\bf s}_{\ell}(-\bar
y)\rho_{j+\ell} (z), \ee in terms of the initial
condition $\rho(z)$. The moments (0.16) give rise to
the polynomials $p_n(z;\bar x,\bar y,\bar t)$, as in
(0.4), which, in turn, give rise to $2m+1$-band
matrices $L$, via $z^m p=L p$. Then $L$ satisfies the
following
 equations\footnote{Note $\overline{L^{1/m}}$ and
 $\underline{L^{1/m}}$ are the {\em
right} $m^{{\rm th}}$ roots and {\em left} $m^{{\rm
th}}$ roots, so that: \bean
\overline{L^{i/m}}&=&(\overline{L^{1/m}})^i
~~\mbox{where}~~\overline{L^{1/m}}=\Lb+\sum_{k\leq
0}b_k\Lb^k \nonumber\\
\underline{L^{i/m}}&=&(\underline{L^{1/m}})^i
 ~~\mbox{where}~~\underline{L^{1/m}}=c_{-1}\Lb^{-1}+
\sum_{k\geq 0}c_k\Lb^k. \eean} in the time parameters
$(\bar x, \bar
 y, \bar t) $,
$$
\frac{\pl L}{\pl x_i}=[(\overline{L^{i/m}})_+,L],\quad
\frac{\pl L}{\pl
y_i}=[(\underline{L^{i/m}})_-,L],\mbox{\,\,for\,\,}i=1,2,...,m\not|
i
$$
\be
\frac{\pl L}{\pl t_{im}}=[(L^i)_+,L],\quad i=1,2,...\,.
\ee

\medbreak

\noindent {\bf Vertex operators}. In order to obtain
 formulae (0.7) and (0.12)
for the weights, we consider two vertex operators, naturally
associated with the integrable system (0.13) for $2m+1$-band
matrices\footnote{$\chi(\lb)=\mbox{diag} ~(\lb^0,\lb,\lb^2,...)$.}
%$\BX_i(\lb)=\BX_i(\bar x,
%\bar y,
%\bar t; \lb)$ will be playing a central role in this work :
\begin{eqnarray}
\BX_1(\lb)&:=&\chi(\lb)e^{\sum_1^{\iy}\bar t_{mi}\lb^{mi}}
e^{-\sum_1^{\iy}
\frac{\lb^{-mi}}{mi}\frac{\pl}{\pl t_{mi}}}e^{\sum_1^{\iy}
 \bar x_i\lb^i}e^{-\sum_1^{\iy} \frac{\lb^{-i}}{i}\frac{\pl}{\pl
\bar x_i}}\nonumber\\
\BX_2(\lb)&:=&\chi(\lb^{-1})e^{-\sum_1^{\iy}\bar t_{mi}\lb^{mi}}
e^{\sum_1^{\iy}\frac{\lb^{-mi}}{mi}\frac{\pl}{\pl \bar t_{mi}}}
e^{\sum_1^{\iy} \bar y_i\lb^i}e^{-\sum_1^{\iy}
\frac{\lb^{-i}}{i}\frac{\pl}{\pl \bar y_i}}\Lb.
 %\nonumber\\
\end{eqnarray}
The vertex operators (0.19) act on vectors of
functions $\tau(\bar x, \bar y, \bar t)=(\tau_n(\bar
x, \bar y, \bar t))_{n\geq 0}$. In \cite{AvM5}, we
showed general linear combinations of them are the
precise implementation of the Darboux transform (0.6)
and (0.9) at the level of $\tau$-functions; see
theorems 4.1 and 4.2. Then in the end, we set $(\bar
x, \bar y, \bar t)=(0,0,0)$, which yield formulae
(0.7) and (0.12) for the new weights.

It is well-known that the vertex operators generate Virasoro-like
symmetries at the level of the $\tau$-functions, which translate
into symmetries at the level of the ``wave"-functions for band
matrices. For the study of such symmetries, see Dickey
 \cite{D1,D3} and
\cite{ASV1}. For an extensive exposition on Darboux transforms, see
the book by Matveev and Salle \cite{MS}.

%\medbreak

%\newpage

\noindent {\bf Examples:}
 {\bf 1. Tridiagonal matrices}: A single
weight leads to a moment matrix $m_{\iy}$ with $\Lb
m_{\iy}=
 m_{\iy}\Lb$ and a
tridiagonal matrix $L$; formulae (0.15) reduce to one
set of times
 $t:=\bar t= (t_1, t_2, \ldots)$. Equations (0.18) become
 the standard Toda lattice, with $\tau$-functions
  \be \tau_n(t)=\det m_n \left(\rho(z) e^{\sum_1^{\iy}
    t_iz^i}\right).\ee
  The
standard Toda lattice vertex operator, introduced by
us in \cite{AvM2} and obtained from (0.19),
 \be
 \BX(t,\lb)=\Lb^{-1}\chi(\lb^2)e^{\sum_1^{\iy}
t_i\lb^i}e^{-2\sum_1^{\iy} \frac{\lb^{-i}}{i}
\frac{\pl}{\pl t_i}}
 \ee
has the surprising property that, given a Toda
$\tau$-vector $\tau(t)=(\tau_0,\tau_1,\ldots)$, the
vector\footnote{For $\al \in \BC$, define $
[\al]=(\frac{\al^{}}{1},\frac{\al^{2}}{2},\frac{\al^{3}}{3},
\ldots)\in \BC^{\iy}$.}
\be
 \tau(t) + c\BX(t,\lb) \tau(t)= \left(
 \tau_n(t) + c\lb^{2n-2}e^{\sum_1^{\iy}
t_i\lb^i} \tau_{n-1}(t-2[\lb^{-1}])\right)_{n \geq 0}
\ee
  is again a Toda
 $\tau$-vector. This precise operation can be
  implemented by a
 UL-Darboux transform, followed by a LU-Darboux
 transform and a limit. Note the UL-Darboux transform (resp.
 LU-Darboux transform) amounts, for a tridiagonal
 matrix, to a factorization of $L-\lb I$ into an upper-
~ times a lower-triangular matrix (resp. lower-
 times an upper-triangular matrix), and to multiplying the
 factors in the opposite order. The vertex operator
 above translates into adding a delta-function to the
 original weight. This establishes a dictionary
 between several points of view (explained in section 6)
 :
$$ L-\lb=L_+ L_- \longmapsto L'-\lb:=L_-
L_+\longmapsto L'-\mu=L'_- L'_+\longmapsto
L''-\mu:=L'_+ L'_-, $$
 $$\left\updownarrow \vbox
to0.5cm{\vss}\right.$$
 $$\rho(z) \longmapsto \rho(z)+c
\dt(\lb-z)$$
 $$ \left\updownarrow \vbox
to0.5cm{\vss}\right.$$
 \be \tau + c\BX \tau\ee

%$2m+1$-step ``classical" polynomials

\noindent {\bf 2. ``Classical" polynomials, satisfying
$2m+1$-step relations}: Given moments $\mu_i:=\la z^i,
\rho_0(z)\ra$, associated with a single weight
$\rho_0$ for standard orthogonal polynomials,
satisfying for fixed integer $m\geq 1$,
 $$ \int_{\BR}\left|z^{j} \rho_0(z)\right|
dz< \iy, ~~j \geq -m+1 ,$$
   we define in section 7 new monic polynomials
 ${\tilde p}_n^{(1)}(z)$,
defined by a new moment matrix $\tilde m_{\iy}$, which
coincides with the old moment matrix
$m_{\iy}=(\mu_{i+j})_{i,j\geq 0}$ associated with the
standard orthogonal polynomials, except for the first
column. The ${\tilde p}_n^{(1)}(z)$, defined by

\medbreak

\noindent$\displaystyle{(\det  {\tilde m}_n)~
  {\tilde p}_n^{(1)}(z)}=$
\hfill
$$
\det\pmatrix{\displaystyle{\sum_{k=0}^{m-1}}\mu_{-k}d_{m-k-1}+
 \,c_0&\mu_1&\mu_2&...&1\cr
 \displaystyle{\sum_{k=0}^{m-1}}\mu_{1-k}d_{m-k-1}
 -c_1&\mu_2&\mu_3&
 ...&z\cr
 \displaystyle{\sum_{k=0}^{m-1}}\mu_{2-k}d_{m-k-1}
 +2!c_2&%
 \mu_3&\mu_4&...&z^{2}\cr %
 \vdots&\vdots&\vdots&...&\vdots \cr
 \displaystyle{\sum_{k=0}^{m-1}}\mu_{m-k-1}d_{m-k-1}
 +(-1)^{m-1}(m-1)!c_{m-1}&%
 \mu_{m}&\mu_{m+1}&...&z^{m-1}\cr
 \displaystyle{\sum_{k=0}^{m-1}}\mu_{m-k}d_{m-k-1}
 &%
 \mu_{m+1}&\mu_{m+2}&...&z^{m}\cr
 \vdots%
 &\vdots&\vdots
 &...&\vdots\cr
 \displaystyle{\sum_{k=0}^{m-1}}\mu_{n-k}d_{m-k-1}
 &%
 \mu_{n+1}&\mu_{n+2}&...&z^{n}}, $$
 satisfy $2m+1$-step relations, i.e.,
 $$
 z^m p^{(1)}(z)= L p^{(1)}(z),~~\mbox{with a $2m+1$-
 band matrix $L$}.
 $$
It remains an interesting open question to find out
whether such polynomials satisfy differential
equations; on such matters, see section 7.

%\newpage

\section{Borel decomposition and the 2-Toda lattice}

\setcounter{equation}{0}

 In \cite{AvM3,AvM4}, we
considered the following differential equations for the bi-infinite
or semi-infinite matrix $m_{\iy}$
\begin{equation}
{\pl m_{\iy}\over\pl x_n}=\Lb^n m_{\iy},\quad {\pl m_{\iy}\over\pl
y_n}=-m_{\iy}\Lb^{\top n},\quad n=1,2,...,
\end{equation}
where the matrix $\Lb=(\dt_{i,j-1})_{i,j\in\BZ}$ is the shift
matrix; then (1.1) has the following solutions:
\begin{equation}
m_{\iy}(x,y)=e^{\sum_1^{\iy} x_n\Lb^n}m_{\iy}(0,0)
 e^{-\sum_1^{\iy} y_n\Lb^{\top
n}}\end{equation} in terms of some initial condition
$m_{\iy}(0,0)$. In this general setup, the matrix
$m_{\iy}$ is a general matrix, thus not necessarily
generated by weights $\rho$.

Consider  the Borel decomposition $m_{\iy}=S_1^{-1}S_2$, for
\begin{eqnarray*}
S_1\in G_-&=&\{\mbox{lower-triangular invertible matrices, with
1's on the diagonal}\}\\
S_2\in G_+&=&\{\mbox{upper-triangular invertible matrices}\},
\end{eqnarray*}
with corresponding Lie algebras $g_-,g_+$; then setting
$\LR_1:=S_1\Lb S^{-1}$,
\begin{eqnarray*}
S_1\frac{\pl m_{\iy}}{\pl x_n}S_2^{-1}&=&S_1\frac{\pl
S_1^{-1}S_2}{\pl x_n}S_2^{-1}=-\frac{\pl S_1 }{\pl
x_n} S^{-1}_1+\frac{\pl S_2}{\pl x_n} S_2^{-1}\in
g_-+g_+\\ &=&S_1\Lb^n m_{\iy}
S_2^{-1}=S_1\Lb^nS_1^{-1}=\LR_1^n=(\LR_1^n)_-+(\LR_1^n)_+\in
g_-+g_+;
\end{eqnarray*}
the uniqueness of the decomposition $g_-+g_+$ leads to
$$
-\frac{\pl S_1}{\pl x_n}S_1^{-1}=(\LR_1^n)_-,\quad\frac{\pl S_2}{\pl
x_n}S_2^{-1}=(\LR_1^n)_+.
$$
Similarly setting $\LR_2=S_2\Lb^{\top}S_2^{-1}$, we find
$$
-\frac{\pl S_1}{\pl y_n}S_1^{-1}=-(\LR_2^n)_-,\quad
\frac{\pl S_2}{\pl y_n}S_2^{-1}=-(\LR_2^n)_+.
$$
This leads to the 2-Toda equations for $S_1,S_2$ and $\LR_1,\LR_2$:
\be
\frac{\pl S_{1,2}}{\pl x_n}=\mp(\LR_1^n)_\mp S_{1,2},\quad
\frac{\pl S_{1,2}}{\pl y_n}=\pm(\LR_2^n)_\mp S_{1,2}
\ee

\be
\frac{\pl \LR_i}{\pl x_n}=[(\LR_1^n)_+,\LR_i],\quad
\frac{\pl \LR_i}{\pl y_n}=[(\LR_2^n)_-,\LR_i],\quad
i=1,2,... \ee By 2-Toda theory \cite{AvM4} the problem
is solved in terms of a sequence of tau-functions
\be
\tau_n(x,y)=\det m_n(x,y),%~~\mbox{for}~ n\in \BZ
\ee
with $m_n(x,y)$ defined below:

\noindent \underline{bi-infinite case} ($n \in \BZ$):
$$m_n(x,y):=\left(\mu_{ij} (x,y)\right)_{-\iy<i,j\leq
n-1},$$

\noindent \underline{semi-infinite case} ($n\geq 0$):
\be
m_n(x,y):=\left(\mu_{ij}(x,y)\right)_{0\leq i,j\leq
n-1},~~\mbox{with}~\tau_0=1.
\ee

The two pairs of wave functions $\Psi=(\Psi_1,\Psi_2)$
and $\Psi^{\ast}=(\Psi_1^{\ast},\Psi_2^{\ast})$
defined by \footnote{In this section, \bean
\chi(z)&=&~\mbox{diag}~(...,z^{-1},z^{0},z^{1},...)
 ~\mbox{in the bi-infinite case}
\\ &=&~\mbox{diag}~(z^{0},z^{1},...)~\mbox{
in the semi-infinite case}.\eean }
 $$ \Psi_{1}(z;x,y)=e^{\sum_1^{\iy} x_i z^{
i}}S_{1}\chi(z),~~\Psi^{\ast}_{1}(z;x,y)
=e^{-\sum_1^{\iy} x_i z^{
i}}\left(S^{\top}_{1}\right)^{-1}\chi(z^{-1}) $$
\be
\Psi_{2}(z;x,y)=e^{\sum_1^{\iy} y_i z^{-
i}}S_{2}\chi(z),~~\Psi^{\ast}_{2}(z;x,y)
= e^{-\sum_1^{\iy} y_i z^{-
i}}\left(S^{\top}_{2}\right)^{-1}\chi(z^{-1})
\ee
 satisfy
 $$
\LR_1\Psi_1=z\Psi_1,~\LR_2\Psi_2=z^{-1}\Psi_2,\quad
 \LR_1^{\top}\Psi_1^{\ast}=z\Psi_1^{\ast},~\LR_2^{\top}
 \Psi_2^{\ast}=z^{-1}\Psi_2^{\ast},
 $$
 and
\be
\left\{
\begin{array}{l}
\frac{\pl }{\pl x_n}\Psi_i=(\LR_1^n)_+\Psi_i\\
\frac{\pl }{\pl y_n}\Psi_i=(\LR_2^n)_-\Psi_i
 \end{array}
\right.~~~~\left\{
\begin{array}{l}
\frac{\pl}{\pl x_n}\Psi_i^* =-((\LR_1^n)_+)^{\top}
\Psi_i^{\ast}\\ \frac{\pl}{\pl y_n}\Psi_i^*
=-((\LR_2^n)_-)^{\top}\Psi_i^*.
\end{array}
\right.
\ee

In \cite{UT}, with a slight notational modification \cite{ASV2},
the wave functions are shown to have the following $\tau$-function
representation:% $\tau(n,t,s)=
%\tau_n(t_1,t_2,\dots;s_1,s_2,\dots),\quad n\in\BZ$,
%to wit:
\bea
\Psi_1(z;x,y)&=&\biggl(
        \frac{\tau_n(x-[z^{-1}],y)}{\tau_n(x,y)}
e^{\sum^{\iy}_1 x_iz^i}z^n
\biggr)_{n\in\BZ} \nonumber\\
\Psi_2(z;x,y)&=&\biggl(
        \frac{\tau_{n+1}(x,y-[z])}{\tau_n(x,y)}
e^{\sum^{\iy}_1 y_iz^{-i}}z^n
\biggr)_{n\in\BZ} \nonumber\\
\Psi^*_1(z;x,y)&=&\Biggl(\frac{\tau_{n+1}(x+[z^{-1}],y)}
{\tau_{n+1}(x,y)}e^{-\sum^{\iy}_1 x_iz^i}z^{-n}\Biggr)_{n\in\BZ}
\nonumber\\
\Psi^*_2(z;x,y)&=&\Biggl(\frac{\tau_n(x,y+[z])}
{\tau_{n+1}(x,y)}e^{-\sum^{\iy}_1 y_i
z^{-i}}z^{-n}\Biggr)_{n\in\BZ},
\eea
with the following bilinear identities satisfied for the wave and
adjoint wave functions $\Psi$ and $\Psi^{\ast}$, for all
$m,n\in\BZ$ (bi-infinite) and $m,n\geq 0$ (semi-infinite) and
$x,y,x',y'\in\BC^{\iy}$:
\begin{equation}
\oint_{z=\iy} \Psi_{1n}(z;x,y)
 \Psi_{1m}^{\ast} (z;x',y')
\frac{dz}{2 \pi
i z}=\oint_{z=0} \Psi_{2n}(z;x,y)
 \Psi_{2m}^{\ast}(z;x',y')
\frac{dz}{2
\pi i z}.
\end{equation}
The $\tau$-functions\footnote{The first contour runs
clockwise about a small neighborhood of $z=\iy$, while the
second runs counter-clockwise about $z=0$.} satisfy the
following bilinear identities:
$$
\oint_{z=\iy}\tau_n(x-[z^{-1}],y)\tau_{m+1}(x'+[z^{-1}],y')
e^{\sum_1^{\iy}(x_i-x'_i)z^i} z^{n-m-1}dz
$$
\be
=\oint_{z=0}\tau_{n+1}(x,y-[z])\tau_m(x',y'+[z])
e^{\sum_1^{\iy}(y_i-y'_i)z^{-i}}z^{n-m-1}dz; \ee they
characterize the 2-Toda lattice $\tau$-functions. Note
(1.7) and (1.9) yield
\be
(S_2)_0=\diag(...,\frac{\tau_{n+1}(x,y)}{\tau_{n}(x,y)},...):=
h(x,y). \ee In \cite{UT}, the facts above are shown
for the bi-infinite case; they can be carefully
specialized to the semi-infinite case, upon setting
$\tau_{-i}=0$ for $i=1,2,...$. \bigbreak

Consider the usual inner-product $\la,\ra$ and an
infinite sequence of weights $\rho(z)=(\rho_0(z),
\rho_1(z),...)$. The moment matrix $m_{\iy}=m_{\iy}(\rho(z))$
 will now depend on $\rho(z)$.  The following proposition will play an
important role in this paper.

\begin{proposition} The
 solution
 %$m_{\iy}=
 %(\mu_{ij})_{0\leq i,j\leq \iy}$ to
 to the equations
\be
{\pl m_{\iy}\over\pl x_n}=\Lb^n m_{\iy},\quad {\pl m_{\iy}\over\pl
y_n}=-m_{\iy}\Lb^{\top n},\quad n=1,2,...,
\ee
 with initial condition
  $$m_{\iy}(\rho (z  ;0,0))=
    (\la z^i, \rho_j(z) \ra)_{0\leq i,j\leq \iy},$$
is given by
\be
m_{\iy}=\left( \la z^i,\rho_j(z;x,y) \ra  \right)
_{i,j\geq 0},
\ee
%for the usual inner product $\la ~,~ \ra$ in $\BR$ and for
%$x,y$-deformations
where the weights $\rho_j(z;x,y)$ evolve as follows\footnote{The elementary Schur polynomials are
defined in footnote 4; also $\frac{\pl {\bf s}_i}{\pl
x_k}={\bf s}_{i-k}$.}
\be
\rho_j(z;x,y)=e^{\sum^{\iy}_1 x_iz^i}
\sum_{\ell=0}^{\iy}{\bf s}_{\ell}(-y)
\rho_{j+\ell}(z), \ee in terms of the initial
condition $\rho(z;0,0)=(\rho_0(z),
\rho_1(z),...)$.
\end{proposition}

\proof
%of a family of weights $\rho_{j}(z)$, $j=0,1,2,...$.
 Indeed, one checks, from (1.15),
$$
 \frac{\pl\rho_{j}}{\pl x_{k}}=z^k\rho_j(z;x,y) \quad
 \frac{\pl\rho_{j}}{\pl y_{k}}=-e^{\sum_1^{\iy} x_{i}z^i}
 \sum_{\ell=k}^{\iy}
{\bf
s}_{\ell-k}(-y)\rho_{j+\ell}(z)=-\rho_{j+k}(z;x,y), $$
from which it follows that
 $$
  \frac{\pl}{\pl
x_{k}}\mu_{ij}(\rho(z;x,y))=\frac{\pl}{\pl x_{k}} \la
z^i,\rho_j (z;x,y) \ra =
 \la z^{i+k}, \rho_{j}(z;x,y)\ra = \mu_{i + k,j}(\rho(z;x,y)),
 $$
$$
\frac{\pl}{\pl y_{k}}\mu_{{ij}}(\rho(z;x,y))=\frac{\pl}{\pl y_{k}} \la
z^i,\rho_j(z:x,y)\ra=-\left\la z^i,
 \rho_{j+k}(z;x,y)\right\ra=-\mu_{i,j + k}(\rho(z;x,y)),
$$
which is equivalent to (1.13). Here is an alternative way of
checking this fact:
 since, from (1.14),
$$
(\Lb^k m_{\iy}(\rho(z;x,y)))_{ij}=\la z^{i+k}, \rho_{j}(z;x,y) \ra ~~
 \mbox{ and }~~
(m_{\iy}(\rho(z;x,y))\Lb^{\top k})_{ij}=\la z^i,\rho_{j+k}(z;x,y)\ra,
$$
one checks

\bigbreak

\noindent$\displaystyle{e^{\sum_{1}^{\iy} x_{n}\Lb^n} \la
z^i,\rho_{j}(z;0,0)\ra_{0\leq i,j \leq
\iy} e^{-\sum_{1}^{\iy}y_{n}\Lb^{\top n}}}
$
\begin{eqnarray}
~~~~~~~~~~ &=& \sum_{k=0}^{\iy}{\bf
s}_{k}(x)\Lb^k\left\la z^{i}, \rho_{j}(z;0,0)
\right\ra_{0\leq i,j < \iy}
 \sum_{\ell=0}^{\iy} {\bf s}_{\ell}(-y)
  \Lb^{\top \ell}\nonumber\\
&=& \sum_{k,\ell=0}^{\iy}{\bf s}_{k}(x)~
 \left\la z^{i+k}, \rho_{j+\ell}(z;0,0)\right\ra
  _{0\leq i,j <\iy} ~
 {\bf s}_{\ell}(-y)\nonumber\\
&=&\left\la e^{\sum_1^{\iy}x_k z^k} z^i,
\sum_{\ell=0}^{\iy}
 {\bf s}_{\ell}(-y)\rho_{j+\ell}(z;0,0)\right
  \ra_{0\leq i,j < \iy}\nonumber\\
 &=& \left\la z^i, \rho_j(z;x,y)     \right
  \ra_{0\leq i,j < \iy}.
\end{eqnarray}
\qed

%\remark We shall denote the sequence of $\tau_n$'s going with this example
%by $\tau_n(\rho)$, where $\rho=(\rho_0,\rho_1,...)$.

%\newpage

\section{Reductions of the 2-Toda Lattice}

\noindent {\bf Reduction from 2-Toda to $2m+1$-band matrices}
:
\bigbreak

For convenience, we define new vectors $\bar x,\bar y,\bar t \in
\BC^{\iy}$, based on the vectors $x,y
\in
\BC^{\iy}$:
\bean
\bar x&=&(x_1,...,x_{m-1},0,x_{m+1},...,
x_{2m-1},0,x_{2m+1},...)
\\
\bar y&=&(y_1,...,y_{m-1},0,y_{m+1},...
,y_{2m-1},0,y_{2m+1},...)\\
\bar t&=&(0,...,0,t_m,0,...,0,t_{2m} ,0,...,0,t_{3m},0,...),
\eean
with
\be
t_{km}:=x_{km}-y_{km}\mbox{ for }k=1,2,...  ~.
\ee
Notice in this subsection, $\LR_1$ and $\LR_2$ are bi-infinite. In
the next subsection, we shall specialize this to the semi-infinite
case.

Recall from section 1,
 $$ m_{\iy}=S_1^{-1}S_2,~ {\cal
L}_1=S_1\Lb S_1^{-1}, ~{\cal L}_2=S_2\Lb^{\top}
S_2^{-1}~\mbox{and}~\tau_n=\det m_n.$$

\begin{proposition}
Whenever $\tau_n(x,y)\neq 0$ for all $n\in \BZ$, the following
three statements are equivalent:
\newline\noindent (i)    $~~~\Lb^m m_{\iy}=m_{\iy}\Lb^{\top m} $
\newline\noindent (ii)  $~\LR^m_1=\LR^m_2$, in which
case
$~~\LR^m_1$ is a $2m+1$-band matrix.
\newline\noindent (iii)$~~\LR_1,~~\LR_2,~m_{\iy}$ and $\tau_n$ are
functions of $\bar x$,$\bar y$ and $\bar t$ only.

\noindent Also (i) or (ii) are invariant manifolds of
the vector fields ${\pl m_{\iy}\over\pl x_n}=\Lb^n
m_{\iy},\quad {\pl m_{\iy}\over\pl
y_n}=-m_{\iy}\Lb^{\top n},\quad n=1,2,..., $.
\end{proposition}

\proof
Indeed, by the invertibility of $S_{1}$ and $S_{2}$
 under the proviso
 above, and remembering the splitting $m_{\iy}=S_1^{-1}S_2$,
  we have that (i) holds if and only if
\be
\LR^m_{1}=S_{1}\Lb^mS_{1}^{-1}=S_{1}\Lb^m
m_{\iy}S_{2}^{-1}=S_{1}m_{\iy}\Lb^{\top m} S_{2}^{-1} =
S_{2}\Lb^{\top m}S_{2}^{-1}=\LR_{2}^m.
\ee
Also note that (i) is equivalent to
$$
0=\Lb^{km} m_{\iy}-m_{\iy}\Lb^{\top km}=
 \left(\frac{\pl}{\pl x_{km}}+
 \frac{\pl}{\pl y_{kn}}\right) m_{\iy}, \quad,k=1,2,\ldots.
$$
This is also tantamount to statement (iii),
 because the invariance
 of $m_{\iy}$ under $\pl/\pl x_{km}+\pl/\pl y_{km}$ implies the
invariance of $\LR_{1}$, $\LR_{2}$ and $\tau_{n}$. From the
solution (1.2), if (i) holds at $(x,y)=(0,0)$, it holds for all
$(x,y)$, and thus, by (2.2), if (ii) holds at $(0,0)$, it also
holds for all $(x,y)$.
\qed

From Proposition 2.1, it follows that the Toda vector fields
respect the band structure of $L:=\LR^m_1=\LR^m_2$, i.e., it is an
invariant manifold of the flow. Therefore the Toda theory can be
recast purely in terms of the $2m+1$-{\em band matrix} of the form
\bea
L&=&\sum_{-m\leq i\leq m}a_i\Lb^i\nonumber\\ &=&
\left(
\begin{array}{lll|llll}
\ddots          &\ddots&     &\ddots & &\ddots &
\mbox{\Huge o}            \\ a_{-m+1}(-1)&...
&a_{0}(-1) &a_1(-1) &...&~~~~1& \\ \hline a_{-m}(0)
&...     &a_{-1}(0)  &a_0(0) &...&a_{m-1}(0) &1\\
~~~\mbox{\Huge o} &\ddots&      &~~~\ddots & &\ddots
&~~~\ddots
\\
\end{array}\right),\nonumber\\
\eea with $a_i$ being diagonal matrices and $a_m=I$.
The vector fields below involve the $i$th powers
$\overline{L^{i/m}}={\cal L}_1^i$ and
$\underline{L^{i/m}}={\cal L}_2^i$ of the {\em right}
$m^{{\rm th}}$ roots $\overline{L^{1/m}}={\cal L}_1$
and {\em left} $m^{{\rm th}}$ roots
$\underline{L^{1/m}}={\cal L}_2$ respectively; see
also footnote 6.

 The
{\em $m$-reduced Toda lattice vector fields} on $L$
are as follows: $$ \frac{\pl L}{\pl
x_i}=[(\overline{L^{i/m}})_+,L],\quad \frac{\pl L}{\pl
y_i}=[(\underline{L^{i/m}})_-,L],\mbox{\,\,for\,\,}i=1,2,...,m\not|
i $$
\be
\frac{\pl L}{\pl t_{im}}=[(L^i)_+,L],\quad
i=1,2,...\,. \ee
 Then $L$ can be expressed in terms of
a string of $\tau$-functions
\be
\tau_n:=\tau_n(\bar x,\bar y,\bar t), \ee which in the
semi-infinite case will take on a very concrete form.

\bigbreak

 \noindent {\bf Reduction from bi-infinite to
semi-infinite 2-Toda } : \bigbreak

In this section we focus on the Borel decomposition of
section 1, but specifically for semi-infinite matrices
$m_{\iy}=(\mu_{ij})_{i,j \geq 0}$, where it is unique.
Remember the decomposition $m_{\iy}=S^{-1}_1S_2$,
where $S_1$ is lower-triangular, with $1$'s on the
diagonal and where $S_2$ is upper-triangular with
$h_n=\det(m_{n+1})/\det(m_n)$ on the diagonal, by
(1.12). Let $h$ denote such a diagonal matrix. For any
matrix $m_{\iy}$, define $\SR(m_{\iy}):=S_1$ and
$h(m_{\iy}):=h$, as functions of the matrix $m_{\iy}$.
Following \cite{AvM3}, we write the Borel
decomposition, as follows
 \be m_{\iy}=S_1^{-1}
S_2=\left(\SR(m_{\iy}\right))^{-1}h(m_{\iy}) \left(
\SR (m_{\iy}^{\top}) \right)^{\top -1}.
 \ee
  It
leads naturally to vectors of monic bi-orthogonal
polynomials
\be
p^{(1)}(z)=\SR(m_{\iy}) \chi(z)= S_1 \chi(z)
~~\mbox{and}~~p^{(2)}(z)=\SR (m_{\iy}^{\top}) \chi(z)
=h (S_2^{\top})^{-1}
\chi(z).
\ee
 Upon introducing a formal inner-product $\la~,~\ra_0$, where
 $\la y^i,z^j\ra_0 =\mu_{ij}$, the polynomials
 $p^{(1)}(z)$ and $p^{(2)}(z)$ enjoy the
following orthogonality property, using (2.6):
\be
\left( \la  p_i^{(1)},p^{(2)}_j \ra_0 \right)_{i,j\geq
0} =S_1 m
\left(h(S_2^{\top})^{-1}\right)^{\top}=\SR(m_{\iy})m_{\iy}
\SR (m_{\iy}^{\top})^{\top}=h. \ee Letting the
semi-infinite matrix $m_{\iy}$ evolve according to the
differential equations (1.1), namely
$$
{\pl m_{\iy}\over\pl x_n}=\Lb^n m_{\iy},\quad {\pl m_{\iy}\over\pl
y_n}=-m_{\iy}\Lb^{\top n},\quad n=1,2,...,
 $$
  we have shown, in
\cite{AvM3}, that the wave functions $\Psi_1$ and
$\Psi_2^{\ast}$ have the following representation in
terms of the bi-orthogonal polynomials constructed
from $m_{\iy}(x,y)$ in (2.7): \bea
\Psi_1(z;x,y)&=&e^{\sum x_kz^k}p^{(1)}(z;x,y)=e^{\sum
x_kz^k}S_1\chi(z)
\\
& & \nonumber\\ \Psi_2^*(z;x,y)&=&e^{-\sum
y_kz^{-k}}h^{-1}p^{(2)}(z^{-1};x,y)= e^{-\sum
y_kz^{-k}}(S_2^{-1})^{\top}\chi(z^{-1}),\nonumber\\
\eea with the $p_n$'s being expressed in terms of
$\tau$-functions $\tau_n$ of 2-Toda:
\be
p_n^{(1)}(z;x,y)=z^n\frac{\tau_n(x-[z^{-1}],y)} {\tau_n(x,y)}~,~~~~
~p_n^{(2)}(z;x,y)=z^n\frac{\tau_n(x,y+[z^{-1}])} {\tau_n(x,y)}.
\ee
and
\be
\tau_n(x,y)=\det m_n (x,y)
~~\mbox{and}~~h_n=\frac{\tau_{n+1}(x,y)}{\tau_{n}(x,y)}
\ee In \cite{AvM3}, we have shown the following matrix
representation for the bi-orthogonal polynomials,
which then leads, using (2.7), to a representation of
the lower-triangular matrices $\SR(m_{\iy})$ and
$\SR(m^{\top}_{\iy})$:
\be
p^{(1)}_n(z;x,y)=\frac{1}{\tau_n(x,y)}\det\left(
\begin{tabular}{lll|l}
$\mu_{00}$&...&$\mu_{0,n-1}$&1\\ $
 \vdots$& &$\vdots$&$\vdots$\\
$\mu_{n-1,0}$&...&$\mu_{n-1,n-1}$&$z^{n-1}$\\
\hline
$\mu_{n,0} $&...&$\mu_{n,n-1}$&$z^n$
\end{tabular}\right)
\ee
\be
p^{(2)}_n(z;x,y)=\frac{1}{\tau_n(x,y)}\det\left(
\begin{tabular}{lll|l}
$\mu_{00}$&...&$\mu_{n-1,0}$&1\\ $\vdots$& &$\vdots$&$
\vdots$\\
$\mu_{0,n-1}$&...&$\mu_{n-1,n-1}$&$z^{n-1}$\\
\hline
$\mu_{0,n} $&...&$\mu_{n-1,n}$&$z^n$
\end{tabular}\right)
\ee

\vspace{1cm}

Assume now the moments $\mu_{ij}$ are given by weights
$\rho(z)=(\rho_0(z),
\rho_1(z),...)$; then
$$
\tau_n(x,y)={\det
\left(\la z^i,\rho_j(z;x,y)    \ra   \right)_{0\leq i,j
\leq n-1}}=D_n( \rho(x,y)),
 $$
where $\rho_j(z;x,y)$ is given by (1.15), i.e.,
$$
 \rho_j(z;x,y)=e^{\sum^{\iy}_1 x_iz^i}
\sum_{\ell=0}^{\iy}{\bf s}_{\ell}(-y)
\rho_{j+\ell}(z).
$$

\begin{lemma}
In the context of Proposition 1.1, the polynomials
above have the following alternative representation in
terms of the entries $\mu_{ij}=\la z^i,
\rho_j(z;x,y)\ra $ of m:
\begin{eqnarray}
p^{(1)}_n(\lb;x,y)&=&\frac{\det\left( \la z^i,(\lb-z)
\rho_j(z;x,y) \ra \right)_{0\leq i,j \leq n-1}} {\det
\left(\la z^i,\rho_j(z;x,y)    \ra   \right)_{0\leq i,j
\leq n-1}}\nonumber\\ &=&\frac{\det(\lb
\mu_{ij}-\mu_{i+1,j})_{0\leq i,j \leq n-1}}{\tau_n(x,y)}\\ &&\nonumber\\
p^{(2)}_n(\lb;x,y)&=&\frac{\det\left( \la z^i,\lb
\rho_j(z;x,y)-\rho_{j+1}(z;x,y) \ra \right)_{0\leq i,j
\leq n-1}} {\det \left(\la z^i, \rho_j(z;x,y)    \ra
\right)_{0\leq i,j \leq n-1}}\nonumber\\
&=&\frac{\det(\lb \mu_{ij}-\mu_{i,j+1})_{0\leq i,j
\leq n-1}}{\tau_n(x,y)}
\end{eqnarray}
\end{lemma}

\proof The proof follows from the representation
(2.11) of $p_n^{(1)}$ above, the representation (1.5)
and (1.6) of $\tau_n$, the representation (1.15) of
$\rho_j$ and from the following identities:
\begin{eqnarray*}
\lb\mu_{ij}(x-[\lb^{-1}],y)
 &:=& \lb\left\la z^i,
 \rho_{j}(z;x-[\lb^{-1}],y)\right\ra\\
&=&\lb \left\la z^i~,~ e^{\sum_{1}^{\iy}
  \left(x_{i}-\frac{\lb^{-i}}{i}\right)z^i}
    \sum_{\ell=0}^{\iy} {\bf s}_{\ell}(-y)
  \rho_{j+\ell}(z) \right\ra\\
&=& \lb\left\la z^i,\left(1-\frac{z}{\lb}\right)
 \rho_{j}(z;x,y)\right\ra\\
&=&\left\la z^i,(\lb-z)\rho_{j}(z;x,y)\right\ra\\
&=&\lb\mu_{ij}(x,y)-\mu_{i+1,j}(x,y),
\end{eqnarray*}
and
\begin{eqnarray*}
\lb\mu_{ij}(x, y+[\lb^{-1}])
 &:=& \lb\left\la z^i,
 \rho_{j}(z;x,y+[\lb^{-1}])\right\ra\\
&=&\lb ~ \left\la z^i, e^{\sum_{1}^{\iy}
  x_{i}z^i} \sum_{\ell=0}^{\iy} {\bf s}_{\ell}(-y-[\lb^{-1}])
  \rho_{j+\ell}(z;0,0) \right\ra\\
&=& \left\la z^i, e^{\sum_{1}^{\iy} x_{i}z^i}
 \sum_{{\ell=0}}^{\iy} \left(\lb
{\bf s}_{{\ell}}(-y)-{\bf s}_{{\ell}-1}(-y)\right)
\rho_{j+\ell}(z;0,0)\right\ra\\
&=&
\lb\mu_{ij}(x,y)-\mu_{i,j+1}(x,y),
\end{eqnarray*}
which is based on the following identity:
\begin{eqnarray*}
\lb \sum_{0}^{\iy} {\bf s}_{n}(-y-[\lb^{-1}])~z^n &=& \lb
e^{-\sum_{1}^{\iy} (y_{i}+\frac{\lb^{-i}}{i}) z^i} \\
 &=& \lb \sum_{0}^{\iy} {\bf s}_{n}(-y) z^n
 \left(1-\frac{z}{\lb}\right)\\
  &=& \sum_{0}^{\iy}\left(\lb
{\bf s}_{n}(-y)-{\bf s}_{n-1}(-y)\right) z^n.
\end{eqnarray*}
\qed

\begin{corollary}
Given weights $\rho_{0}, \rho_{1}, \ldots, \rho_{n-1}$, the following
identity holds:
$$
\det(\la z^i, (\lb-z)\rho_{j}(z)\ra)_{0\leq i, j\leq n-1}=
\det
\left(
\begin{array}{ccc|c}
 \la z^{0},\rho_{0}(z)\ra  &  \cdots  &
 \la z^{0}, \rho_{{n-1}}(z)\ra  & 1\\
 \vdots  &  & \vdots & \vdots\\
 \la z^n,\rho_{0}(z)\ra  &  \cdots  &
  \la z^n,\rho_{{n-1}}(z)\ra  &  \lb^n \\
\end{array}
\right)
$$
\end{corollary}

\proof  From Lemma 2.2, it follows that $p_{n}^{(1)}$
has two alternative expressions (2.13) and (2.15).
Equating the two leads to the identity above. \qed

\remark Formula (2.15) and hence (2.13) just depend on
the first formula (2.11)
 and $\tau_n=\det(\mu_{ij})_{0\leq i,j \leq n-1}$, with
 $\mu_{ij}(x,y)=\la z^i, e^{\sum x_i z^i} \rho_j(y,t)\ra$.
 The $y$-dependence is unimportant.

\section{From $m$-periodic weight sequences to $2m+1$-band
matrices}

Given the $m$-periodic sequence of weights
\be
\rho=(\rho_{j})_{j\geq 0}=(\rho_{0},\rho_{1},\ldots,
\rho_{{n-1}},z^m\rho_{0}, \ldots,z^m\rho_{m-1}, z^{2m}\rho_{0},\ldots,
z^{2m}\rho_{m-1},\ldots),
\ee
consider the initial value problem
\be
 \frac{\pl m_{\iy}}{\pl x_{n}}=\Lb^n m_{\iy},
 \frac{\pl m_{\iy}}{\pl y_{n}}=
 -m_{\iy} \Lb^{\top n},
 \mbox{ with initial }m_{\iy}(0,0)=(\la
 z_{i},\rho_{j}\ra)_{0\leq i, j<\iy},
\ee
and the associated 2-Toda lattice equations
\be
\frac{\pl \LR_{i}}{\pl x_{n}} = [(\LR_{1}^n)_{+}, \LR_{i}], \quad
\frac{\pl \LR_{i}}{\pl y_{n}} = [(\LR_{2}^n)_{-}, \LR_{i}]
.\ee

In proposition 1.1, we gave the solution to the initial value
problem (3.2) in general, whereas in theorem 3.1, we shall give the
solution for $m$-periodic sequences of weights. This
extra-structure will be important, when we deal with Darboux
transforms.
\begin{theorem}
Given the initial $m$-periodic weights (3.1), the
systems of differential equations (3.2) has the
following solutions with regard to the time parameters
$(\bar x,\bar y, \bar t)$, introduced in (2.1): \bea
m_{\iy}\left(\rho(z;\bar x,\bar y, \bar t)\right)
 &=& \left(\left\la z^i, \rho_{j} (z; \bar x, \bar
 y, \bar t)\right\ra\right)_{0\leq i, j<\iy},
\eea
where
\be
\rho_{j}(z;\bar x,\bar y,\bar t):=
 e^{\sum_1^{\iy} \bar x_{r}z^r}
 e^{\sum_{\ell=1}^{\iy} \bar t_{{\ell m}}z^{\ell m}}
\sum_{\ell=0}^{\iy} {\bf s}_{\ell}(-\bar
y)\rho_{j+\ell} (z). \ee
 is an $m$-periodic sequence of weights. Then the polynomials
$p_n^{(1)}$,
 with $\mu_{ij}:=\mu_{ij}(\rho(z;\bar
 x,\bar y, \bar t))$ and $\tau_n(\bar
 x,\bar y, \bar t)=\det m_n(\rho(z;\bar
 x,\bar y, \bar t))$,
\begin{eqnarray*}
 p^{(1)}_n(z;\bar x,\bar y, \bar t)&=&\frac{1}{\tau_n(\bar
 x,\bar y, \bar t)}\det\left(
 \begin{tabular}{lll|l}
$\mu_{00}$&...&$\mu_{0,n-1}$&1\\ $\vdots$& &$\vdots$&$\vdots$\\
$\mu_{n-1,0}$&...&$\mu_{n-1,n-1}$&$z^{n-1}$\\
\hline
$\mu_{n0}$&...&$\mu_{n,n-1}$&$z^n$
\end{tabular}\right)\\
&=&\frac{\det(z \mu_{ij}-
 \mu_{i+i,j})_{0\leq i,j \leq n-1}}{\tau_n(\bar
 x,\bar y, \bar t)}
\end{eqnarray*}
give rise to matrices $L=\LR_1^m$, defined by $z^m
p^{(1)}=L p^{(1)}$, such that $L=\LR_1^m$ is a
$2m+1$-band matrix. The matrix $\LR_1$ satisfies
equations (3.3) and the $2m+1$-band matrix $L$ the
$m$-reduced Toda lattice equations (2.4).
\end{theorem}

\proof
Since
$$
\rho_{j+km}=z^{km}\rho_{j} ,\quad j,k=0,1,2,\ldots,
$$
 we have
\begin{eqnarray*}
0&=& \la z^i, z^{km} \rho_{j}-\rho_{j+km}\ra
 \\ &=& \la
 z^{i+km},\rho_{j}\ra -\la z^{i}, \rho_{j+km}\ra
 \\ &=&
 \mu_{i+km,j}-\mu_{i,j+km}
 \\ &=&
 \left(\Lb^{km} m_{\iy}-m_{\iy}\Lb^{\top km}\right)_{ij},
 \\
 %&=&
 %\left(\left(\frac{\pl}{\pl x_{km}}+\frac{\pl}{\pl
 %y_{xm}}\right) m_{{\iy}}\right)_{ij},
\end{eqnarray*}
and so $m_{\iy}$ satisfies (i) of proposition 2.1 at $(x,y)=(0,0)$
and hence for all $(x,y)$. Therefore, by proposition 2.1,
$L:=\LR_{1}^m$ is a $2m+1$ band matrix.

From proposition 1.1, we know that the expression below for
$m_{{\iy}}$ is a solution of the initial value problem (3.2). The
proof of (3.4) follows the lines of calculation (1.16).
 From there one
computes

\medbreak

\noindent $m_{\iy}(\rho(z;x,y))$
\begin{eqnarray*}
  &=&e^{\sum_{1}^{\iy} x_{n}\Lb^n} m_{\iy}
  (\rho(z;0,0)) e^{-\sum_{1}^{\iy} y_{n}\Lb^{\top n}}
 \\ &=& e^{\sum_{1}^{\iy} x_{n}\Lb^n}\left\la z^i,
 \rho_{j}(z;0,0)\right\ra
 _{0\leq i,j<\iy} e^{-\sum_{k=1}^{\iy}
 y_{km}\Lb^{\top km}} e^{-\sum_1^{\iy} \bar y_{r} \Lb^{\top r}}
 \\
 &=& \sum^{\iy}_0 {\bf s}_n(x) \Lb^n \left\la z^i, \rho_j(z;0,0)
  \right\ra \sum_0^{\iy} {\bf s}_r(-y_m,-y_{2m},...)
   \Lb^{\top mr}
 e^{-\sum_1^{\iy} \bar y_{r} \Lb^{\top r}}\\
 &=&  \left\la  \sum_0^{\iy} {\bf s}_n(x) z^{i+n},
  \sum_0^{\iy}
 {\bf s}_r(-y_m,-y_{2m},...)\rho_{j+rm}(z;0,0)
   \right\ra_{0\leq i,j<\iy} e^{-\sum_1^{\iy} \bar y_{r}
    \Lb^{\top r}} \\
   &=&  \left\la  e^{\sum_{1}^{\iy} x_r z^r} z^{i},
    \sum_0^{\iy}
 {\bf s}_r(-y_m,-y_{2m},...)z^{rm} \rho_{j}(z;0,0)
   \right\ra_{0\leq i,j<\iy} e^{-\sum_{1}^{\iy} \bar y_{r}
    \Lb^{\top r}} \\
 &=&
 \left\la z^i, e^{\sum_{1}^{\iy} \bar x_{r} z^r}
  e^{\sum_{n=1}^{\iy}
 x_{km}z^{km}} e^{-\sum_{k=1}^{\iy} y_{km}z^{km}} \rho_{j}(z;0,0)
 \right\ra_{0\leq i,j<\iy}
 e^{-\sum_{1}^{\iy}\bar y_{r}\Lb^{\top r}}
 \\ &=&
 \left\la z^i, e^{\sum_{1}^{\iy} \bar x_{r} z^r}
  e^{\sum_{k=1}^{\iy} \bar t_{km}z^{km}}
 \rho_{j}(z;0,0)\right\ra_{0\leq i,j<\iy}
 e^{-\sum_{1}^{\iy}\bar y_{r}\Lb^{\top r}}
 \\ &=&
 \left\la  z^i, e^{\sum_{1}^{\iy} \bar
 x_{r} z^r} e^{\sum_{k=1}^{\iy} \bar t_{km}z^{km}}
 \rho_{j}(z;0,0)\right\ra_{0\leq i,j<\iy}
  \sum_{\ell=0}^{\iy}
 {\bf s}_{\ell}(-\bar y)\Lb^{\top \ell}\\ &=&
   \left\la z^i, e^{\sum_{1}^{\iy} \bar
 x_{r} z^r} e^{\sum_{k=0}^{\iy} \bar t_{km}z^{km}}
 \sum_{\ell=0}^{\iy} {\bf s}_{\ell}(-\bar y)\rho_{j+\ell}(z; 0,0)
 \right\ra_{0\leq i, j<\iy},
\end{eqnarray*}
which establishes (3.4).  The rest follows from (2.13)
(see the last remark of section 2) and Lemma 2.2.
\qed

%\newpage

In the following we show that $m$-periodic sequences
of weights lead to $2m+1$ band matrices, using a
direct proof, thus without invoking the matrices
$\LR_1$ and $\LR_2$ of 2-Toda theory, as in Theorem
3.1. Furthermore, it will be shown that the
polynomials $p_n^{(1)}$ are ``{\em orthogonal}" in the
sense (3.7).
Consider here the slightly more general definition of
$m$-periodic sequences (in comparison to (0.1)):
  \begin{definition} {\em
Generalized
 $m$-periodic sequences of weights} $\rho_i$ satisfy the following condition:
  for $j=0,1,2,\ldots,$
\be
 z^m\rho_j\in\span\{\rho_0,\ldots,\rho_{m+j}\}
 ~~\mbox{and}~
z^m\rho_j(z)=c_{j,m+j}\rho_{m+j}(z)+\ldots
,~\mbox{with}~ c_{j,m+j}\neq 0. \ee \end{definition}
\begin{proposition}
Given a sequence of weights $\rho_{0}(z),\rho_1(z),\ldots$, the monic
polynomials $p_0(z),p_1(z),\ldots,p_j(z),\ldots$ of degree
$0,1,2,\ldots$, defined by
\be
\la p_i(z),\rho_j(z)\ra =0,\quad 0\leq j\leq i-1 \ee
are given by the same formula, as in Theorem 3.1,
namely
\be
 p_n(z)=\frac{1}{\det m_n}\det\left(
 \begin{tabular}{lll|l}
$\mu_{00}$&...&$\mu_{0,n-1}$&$1$\\ $\vdots$&
&$\vdots$&$\vdots$\\
$\mu_{n-1,0}$&...&$\mu_{n-1,n-1}$&$z^{n-1}$\nonumber\\
\hline $\mu_{n0}$&...&$\mu_{n,n-1}$&$z^n$
\end{tabular}\right) %\nonumber\\\nonumber
\ee
 with $\mu_{ij}=\la z^i,\rho_j(z)\ra$, $m_n=\det(\mu_{ij})_{0\leq i,j\leq
n-1}$. Moreover, if the $\rho_i$ are generalized
 $m$-periodic,
then the polynomials (3.7) satisfy a $2m+1$-step
relation; i.e., for
$p(z)=(p_0(z),p_1(z),\ldots)^{\top},$
\be
 z^mp(z)=Lp(z)
\ee
defines a $2m+1$ band matrix $L$, with
$m$ bands above and $m$ below
the diagonal.

\end{proposition}

\proof For $0\leq k\leq n-1$, the inner-product of
$p_n(z)$, given by the right hand side of (3.8),  with
$\rho_k(z)$ automatically vanishes:
 $$(\det m_n)~\la p_n(z),\rho_k(z)\ra
=\det\left(\la
\mu_{i0},\mu_{i1},\ldots,\mu_{ik},\ldots,
\mu_{i,n-1},\mu_{ik}\ra_{i=0,...,n}\right)=0. $$
Furthermore, the orthogonality relation (3.7)
determines the monic $p_n$'s uniquely. To prove the
second assertion, that $L$ is a $2m+1$ band matrix, we
proceed as follows:
 since $z^m\rho_j(z)=
\displaystyle{\sum_{r=0}^{m+j}}c_{jr}\rho_r(z)$,
$j=0,1,\ldots$, we have
\bean
0&=&\left\la z^i,z^m\rho_j-\sum_{r=0}^{m+j}c_{jr}
 \rho_r(z)\right\ra,\qquad \mbox{for
all~}i,j\geq 0,\\
&=&\la z^{i+m},\rho_j\ra - \sum_{r=0}^{m+j}c_{jr}
 \la z^i,\rho_r(z)\ra \\
&=& \mu_{m+i,j}-\sum_{r=0}^{m+j}c_{jr}\mu_{ir},\\
%&=& \mu_{m+i,j}-\sum_{r=0}^{n-1}c_{jr}\mu_{ir}, \qquad
%\mbox{if~}0\leq j\leq n-m-1,
\eean
implying for all $j\geq 0$,
$$
\left(\begin{array}{l}
\mu_{m,j}\\
 \mu_{m+1,j}\\
 \vdots\\
 \mu_{m+n,j}
 \end{array}
 \right)
 =\sum_{r=0}^{m+j}
  c_{jr}
  \left(\begin{array}{l}
\mu_{0,r}\\
 \mu_{1,r}\\
 \vdots\\
 \mu_{n,r}
 \end{array}
 \right).
 $$
Therefore, by (3.8) the following determinant vanishes
for arbitrary $n\geq 0$, as long as $n-1 \geq m+j$,
 $$
%
%&=&\frac{1}{a_{nn}}\sum^n_{i=0}a_{ni}\la z^i,z^m\rho_j(z)\ra \\
0=\frac{1}{D_n(\rho)}\det\left(\begin{array}{lll|l}
\mu_{00}&\ldots&\mu_{0,n-1}&\mu_{mj}\\ \vdots& &\vdots
&\vdots\\
 & & & \\
\hline
\mu_{n,0}&\ldots&\mu_{n,n-1}&\mu_{m+n,j}
\end{array}\right)
 =\la z^m p_n(z),\rho_j(z)\ra ,$$
$\mbox{~for all $j$ such that~}0\leq j\leq n-m-1.$
This implies that
\bean
z^mp_n(z)
&\in &\{\mbox{polynomials~}q(z)~\Big| ~ \la q(z),\rho_j(z)\ra
  =0,~ \mbox{for}~0\leq j\leq n-m-1\},\\
  &=&\span\{p_{n-m}(z),p_{n-m+1}(z),\ldots,\}\\
  &=&\span\{p_{n-m}(z),p_{n-m+1}(z),\ldots,p_{n+m}(z)\}
;\eean the latter identity is valid, because
$z^mp_n(z)$ has degree $n+m$. Therefore $L$ defined by
(3.9) is $2m+1$-band, as claimed, ending the proof of
Proposition 3.3.\qed

\remark    A generalized $m$-periodic sequence of
weights can be transformed in an $m$-periodic sequence
of weights, via an invertible lower-triangular
transformation of the $\rho_i$ in the sequence
$\rho(z)=\left(\rho_j(z)\right)_{j\geq 0}$; the new
sequence of weights thus obtained become $m$-periodic,
i.e., \be z^m \rho_j =z^m \rho_{m+j}. \ee
 Such a transformation leaves the associated
 polynomials (3.8)
 unaffected, as is seen from column operations
 in the defining ratio of determinants in (3.8).
  These polynomials then lead to $2m + 1 $
band matrices $L$, which are thus unaffected by the
lower-triangular operations of the $\rho_i$.

\section{Darboux transformations on $2m+1$-band matrices}

The vertex operators $\BX_i(\lb):=\BX_i(\bar x, \bar y,
 \bar t; \lb)$, introduced in the introduction (see \cite{AvM5}), play a central
 role in this work\footnote{$\chi(\lb)=
 \mbox{diag} (\lb^0,\lb^1,\lb^2,\ldots)$} :
\begin{eqnarray}
\BX_1(\lb)&:=&\chi(\lb)e^{\sum_1^{\iy}\bar t_{mi}\lb^{mi}}
e^{-\sum_1^{\iy}
\frac{\lb^{-mi}}{mi}\frac{\pl}{\pl \bar t_{mi}}}e^{\sum_1^{\iy}
 \bar x_i\lb^i}e^{-\sum_1^{\iy} \frac{\lb^{-i}}{i}\frac{\pl}{\pl
\bar x_i}}\nonumber\\
\BX_2(\lb)&:=&\chi(\lb^{-1})
 e^{-\sum_1^{\iy}\bar t_{mi}\lb^{mi}}
e^{\sum_1^{\iy}\frac{\lb^{-mi}}{mi}\frac{\pl}{\pl \bar t_{mi}}}
e^{\sum_1^{\iy} \bar y_i\lb^i}e^{-\sum_1^{\iy}
\frac{\lb^{-i}}{i}\frac{\pl}{\pl \bar y_i}}\Lb;
%\nonumber\\
\end{eqnarray}
e.g., $\BX_2(\lb)$ acts on the vector $\tau(\bar x,
\bar y, \bar t)$, as follows
$$
\left(\BX_2(\lb) \tau(\bar x, \bar y,\bar t)\right)_n
 =e^{-\sum_1^{\iy}\bar t_{mi}\lb^{mi}}
 e^{\sum_1^{\iy} \bar y_i \lb^i}  \lb^{-n}
 \tau_{n+1}(\bar x, \bar y -[\lb^{-1}],
  \bar t -[\lb^{-1}])
 $$
 where
\bean
 \bar y-[\lb^{-1}]&:=&(y_1-\frac{\lb^{-1}}{1},...,y_{m-1}
 -\frac{\lb^{-(m-1)}}{m-1},0,y_{m+1}-\frac{\lb^{-(m+1)}}{m+1},...
%,y_{2m-1},0,y_{2m+1},
\ldots)\\
\bar t -[\lb^{-1}]
&:=&(0,...,0,t_m -\frac{\lb^{-m}}{m},0,...,0,
 t_{2m}-\frac{\lb^{-2m}}{2m} ,0,...,0,...).
\eean

The following two theorems were established in
\cite{AvM5}, and will be applied in section 5 to the
concrete $\tau_n$'s given by $\tau_n=\det m_n(\rho)$,
with the $\rho_n$'s as in (3.5).

\begin{theorem} ({\bf LU-Darboux transform}) Given the Toda lattice
on semi-infinite $2m+1$-band matrices, each vector
$\Phi(\lb)$ in the $m$-dimensional null-space,
i.e.\footnote{$\om$ is a primitive $m$th root of
unity.}, $$%\mbox{each~}~
\Phi(\lb)=\frac{\tilde\tau}{\tau}
:=\frac{\displaystyle{\sum_{k=0}^{m-1}}
\left(a_k\BX_1(\om^k\lb)\right)\tau} {\tau}\in
(L(t)-\lb^m I)^{-1}(0,0,...) $$ satisfies, as a
function of $\bar x, \bar y,
 \bar t$, the
following equations:
 $$ L\Phi=\lb^m\Phi $$

 \be \frac{\pl\Phi}{\pl
x_i}=(\overline{L^{i/m}})_+\Phi,\quad\quad
\frac{\pl\Phi}{\pl
y_i}=(\underline{L^{i/m}})_-\Phi,\quad\quad
 \frac{\pl\Phi}{\pl t_{ip}}=(L^i)_+\Phi,
  \ee
for $i=1,2,...$ not multiples of $m$ for the $x_i$ and
$y_i$ equations. Each $\Phi(\lb)$ determines an
LU-Darboux transform, depending projectively on the
$m-1$ parameters $a_i$; namely $$ L-\lb^m I\longmapsto
\tilde L-\lb^m I:=(\beta\Lb^0+\Lb)(L-\lb^m
I)(\beta\Lb^0+\Lb)^{-1} $$ with
 \be
  \beta_n=-\frac{\Phi_{n+1}(\lb)}{\Phi_n(\lb)};
  \ee
it acts on $\tau$ as
\be
\tau\longmapsto\tilde\tau=\tau\Phi=
\displaystyle{\sum_{k=0}^{m-1}}
\left(a_k\BX_1(\om^k\lb)\right)
\tau.
\ee
\end{theorem}

Defining $e_i:=(0,...,0,\underbrace{1}_i,0,...)\in
\BR^{\iy}$, as before, we have:
\begin{theorem} ({\bf UL-Darboux transform}) Given the Toda lattice on
semi-infinite $2m+1$-band matrices, the space
$(L-\lb^mI)^{-1}\span\{e_0,e_1,...,e_m\}$ is
$2m$-dimensional and thus depends projectively on
$2m-1$ free parameters, i.e.,
\begin{eqnarray*}
%\mbox{each}~~
\Phi(\lb)=\frac{\Lb\tilde\tau}{\tau}
&:=&\frac{\displaystyle{\sum_{k=0}^{m-1}}\left(a_k\BX_1(\om^k\lb)+b_k
e^{\sum_{1}^{\iy}t_{im}\lb^{im}}\BX_2(\om^k\lb)\right)\tau}
{\tau}
% \in (L(t)-\lb^m I)^{-1}\span
%\{e_0,e_1,...,e_m\}
\\
&&\\
 &\in &(L(t)-\lb^m I)^{-1}\span
\{e_0,e_1,...,e_m\}.
%{\underbrace{ *,*,...,*}_{m}},0,0,...)
\end{eqnarray*}
The vector  $\Phi(\lb)$, as a function of $\bar x,
\bar y,
 \bar t$~, satisfies the same
equations (4.2) and determines a UL-Darboux transform,
with the same $\beta$ as (4.3) (but depending
projectively on $2m-1$ free parameters): $$ L-\lb^m
I\longmapsto \tilde L-\lb^m
I:=(\Lb^{-1}\beta+I)(L-\lb^m I)(\Lb^{-1}\beta+I)^{-1};
$$ it induces a map on $\tau$ : $$
\tau\longmapsto\tilde\tau=\Lb^{-1}(\tau\Phi)= \Lb^{-1}
\sum_{k=0}^{m-1}\left(a_k\BX_1(\om^k\lb)+b_k
e^{\sum_{1}^{\iy}t_{im}\lb^{im}}\BX_2(\om^k\lb)\right)
 \tau
. $$
\end{theorem}

\section{Proof of Theorems 0.1 and 0.2: Induced Darboux maps on $m$-periodic weights}

In order to prove Theorems 0.1 and 0.2, we apply
Theorems 4.1 and 4.2 to the $\tau$-functions given by
$$\tau_n(\bar x,\bar y,\bar t)=D_n(\rho(z;\bar x,\bar
y,\bar t)):=D_n(\rho_0 (z;\bar x,\bar y,\bar t)
 ,\rho_1 (z;\bar x,\bar y,\bar t),\ldots)=\det m_n(\rho(z;\bar x,\bar y,\bar t))$$
with
\be
\rho_{j}(z;\bar x,\bar y,\bar t)=
 e^{\sum_{1}^{\iy} \bar x_{r}z^r}
 e^{\sum_{\ell=1}^{\iy} \bar t_{{\ell m}}z^{\ell m}}
\sum_{\ell=0}^{\iy} {\bf s}_{\ell}(-\bar
y)\rho_{j+\ell} (z), \ee as in (3.5), where the
initial condition
$\rho(z)=\left(\rho_j(z)\right)_{j\geq 0}$ forms an
$m$-periodic sequence of weights. We now perform
Darboux transformations on
 $L (\bar x, \bar y, \bar
 t)$, which satisfies the {\em $m$-reduced Toda lattice
 equations}  (2.4). Then, in the end, put
 $\bar x=\bar y=\bar t=0$.
Theorems 5.1 and 5.2 are the precise analogues of
Theorems 4.1 and 4.2:

 \begin{theorem} ({\bf LU-Darboux})
 The Darboux transform for a semi-infinite $2m + 1$-band
 matrix, generated by the $m$-periodic sequences of weights $\rho(z;\bar x,\bar y,\bar
 t)$ above,
 \be
 L - \lb^m I \mapsto \tilde L - \lb^m I =
  (\beta  \Lb^0 + \Lb) (L -
 \lb^m I) (\beta \Lb^{0} + \Lb)^{-1},
 \ee
 defines a new $2m+1$-band matrix $\tilde L$,
 provided ($\om$ is a primitive $m$th root of unity)
\be
 \beta_n = - \frac{\Phi_{n+1} (\lb)}{\Phi_n (\lb)} ,
 ~~\Phi_n (\lb) =
 \frac{\sum^{m-1}_{k=0} a_k \BX_1 (w^k \lb)
 D_n(\rho(z;\bar x,\bar y,\bar t))}
 {D_n(\rho(z;\bar x,\bar y,\bar t))}
 \ee
\underline{\bf Case 1}  For the special choice $$
\Phi^{(k)}_n (\lb)
=
 a_k\frac{ \BX_1 (w^k \lb)
 D_n(\rho(z;\bar x,\bar y,\bar t))}
 {D_n(\rho(z;\bar x,\bar y,\bar t))}
 $$
  with arbitrary, but fixed $1\leq k\leq n$,
   the Darboux transformation maps
   $\tau_n(\bar x,\bar y,\bar t)=D_n(\rho)$ into a
   $D_n$ associated with a new $m$-periodic sequence
   of weights:
 \begin{eqnarray}
 D_n(\rho(z;\bar x,\bar y,\bar t)) \mapsto \tilde D_n
 &=& D_n(\rho(z;\bar x,\bar y,\bar t)) \Phi^{(k)}_{n} (\lb)\nonumber\\
 &=& %\sum^{m-1}_{k:0}
  \tilde a_{k} D_n
 ((\om^k \lb - z) \rho
  (z;\bar x,\bar y,\bar t) )
 \end{eqnarray}
\underline{\bf Case 2} A general linear combination
 \be
  \Phi_n (\lb) =\frac{\sum^{m-1}_{k=0} a_k \BX_1 (w^k \lb)
 D_n(\rho(z;\bar x,\bar y,\bar t))}
 {D_n(\rho(z;\bar x,\bar y,\bar t))}
  \ee
  leads to the map
 \begin{eqnarray}
\tau_n( \bar x,\bar y,\bar t)=D_n(\rho(z;\bar x,\bar
y,\bar t) \mapsto \tilde \tau_n (\bar x,\bar y,\bar t)
 &=& D_n\left(\rho(z;\bar x,\bar y,\bar t)\right) \Phi^{(k)}_{n} (\lb)\nonumber\\
 &=& \sum^{m-1}_{k=0}
  \tilde a_{k} D_n
 \Bigl((\om^k \lb - z) \rho
  (z;\bar x,\bar y,\bar t) \Bigr)\nonumber\\
&=& (-1)^n \det \left(\bigl\la z^i ,\tilde \rho_0
 \bigr\ra , \la z^i ,\tilde\rho_1 \ra ,\ldots,
 \la z^i ,\tilde\rho_{n} \ra
  \right)_{0 \leq i \leq n},\nonumber\\
 \end{eqnarray}
 where
  \bea
 \tilde \rho_0 :&=&
 \sum_{k=0}^{m-1}
  \tilde a_k \dt (z - \om^k \lb)  \nonumber\\
 \tilde \rho_{\ell} :&=& \rho_{\ell-1} (z ;\bar x, \bar y, \bar t),
 ~~~\mbox{ for } \ell \geq 1,
 \eea
and %$\rho_j = \rho_j (z;\bar x,\bar y,\bar t)$ given as in (3.5) and with
 $$
 \tilde a_k = a_k e^{\sum_{i=1}^{\infty} \bar t_{im} \lb^{im}}
 e^{\sum_{i=1}^{\infty} \bar x_i (\om^k \lb)^i }.
 $$

  \end{theorem}

 \remark For the general case (case 2), (5.6) is the
  determinant a $(n+1)\times (n+1)$
 matrix, instead of $n \times n$.  Therefore, to the
 best of our knowledge,
   this $\tau$-function is not generated in the
    usual way, as a determinant of the $n\times n$
 upper-left corner of the moment matrix.
 If all but one of the $a_k$'s vanish, as in
 case 1,
  then the
 $\tau$-functions are generated in the usual way,
 as appears immediately from the second identity of
 (5.4).
In the next statement, this problem will be absent.
%\newpage

 \begin{theorem} ({\bf UL - Darboux})
 The Darboux transform for a semi-infinite
 $2m + 1$-band matrix, arising from a $m$-periodic
 weights $\rho(z;\bar x,\bar y,\bar
 t)$,
 \be
 L - \lb^m I \longmapsto \tilde L - \lb^m I = (\Lb^{\top} \beta + I)
 (L - \lb^m I) (\Lb^{\top} \beta + I)^{-1} ,
 \ee
maps $L$ into a new $2m + 1$-band matrix $\tilde L$,
provided
 (with $D(\rho):=(D_0(\rho),D_1(\rho),\ldots)$),
  \be
 \beta_n = - \frac{\Phi_{n+1} (\lb)}{\Phi_n (\lb)} ,
 ~~~~\Phi_n (\lb) =
 \frac{\left(\sum_{k=0}^{m-1} \left(a_k \BX_1
  (\om^k \lb) +
 b_k e^{\sum \bar t_{im}
 \lb^{im}} \BX_2 (\om^k \lb)\right) D(\rho)\right)_n}
 {D_n(\rho)}.
 \ee
 It acts on $\tau_n(
 \bar x, \bar y, \bar t)=D_n(\rho(z ;
 \bar x, \bar y, \bar t))$ as follows
\begin{eqnarray*}
 \tau_n:=D_n(\rho(z ;
 \bar x, \bar y, \bar t))
 % =\det \left\la z^i,\rho_j
 %\right\ra_{0\leq i,j\leq n}
 \mapsto \tilde\tau_n &=& D_{n-1}(\rho(z ;
 \bar x, \bar y, \bar t))\Phi_{n-1}(\lb)
 %(\Lb^{-1} \tau \Phi (\lb))_n
 \\
 &=& (-1)^{n-1} \det
  \left(\la z^i , \tilde \rho_0 \ra, \la z^i, \tilde
 \rho_{1} \ra ,\ldots \la z^i ,
 \tilde \rho_{n-1} \ra \right)_{0 \leq i \leq n-1}.
 \end{eqnarray*}
 with
 \bea
 \tilde \rho_0 :&=& \tilde \rho_0 (z ;
 \bar x, \bar y, \bar t) :=
 \sum_{k=0}^{m-1} \left(\tilde a_k \dt
  (z - \om^k \lb) +
 \tilde b_k \frac{\rho_k (z; \bar x,\bar y, \bar t)}
 {z^m - \lb^m }\right)\nonumber\\
 \tilde \rho_{\ell} :&=& \rho_{\ell-1} (z ;\bar x, \bar y, \bar t),
 ~~~\mbox{ for } \ell \geq 1,
 \eea
 where
 \be
 \tilde a_k = a_k e^{\sum_1^{\infty} \bar x_i (\om^k\lb)^i }
 e^{\sum_1^{\infty} \bar t_{im} \lb^{im}} ,~~
 \tilde b_k = - \lb^{m-k} \sum_{j=0}^{m-1} b_j e^{\sum_{i\geq0} \bar y_i
 (\om^j \lb )^i} \om^{-jk}.
 \ee
If $\tilde b_{m-1} \neq 0$, then the
  $\tilde \rho_0,\tilde \rho_1,\ldots $
form a generalized $m$-periodic sequence.
 \end{theorem}

\remark Although the new sequence $\tilde \rho(z;\bar
x,\bar y, \bar t)$ is generalized m-periodic in the
sense of (3.6), it does not lead to a solution
$m_{\iy}$ of the differential equations (3.2); in
other words, it only satisfies (3.5) in the $\bar x$
and $\bar t$ variables, but not in the $\bar y$
variable. Of course, the matrix $\tilde L$ remains a
$2m+1$-band matrix, since it is effectively
constructed from the new polynomials $p_n(z;\bar
x,\bar y, \bar t)$, defined by (3.8) with the new
$\rho$'s; see remark at the end of section 3.

 \begin{corollary}
 An appropriate choice of $a_k $, and appropriate limits $b_k \mapsto
 \infty$ and $\lb \mapsto 0$ in Theorem 5.2 yield
  the following Darboux
 transformation on the weights $\rho(z ; \bar x; \bar y; \bar
 t)$:
 $$\rho=( \rho_0 ,\rho_1 , \rho_2 ,\ldots) \mapsto \tilde\rho=(\tilde \rho_0 ,\tilde\rho_1 ,\tilde \rho_2 ,\ldots),$$
 where
 \bea
 \tilde \rho_0  &=&
 \sum_{k=0}^{m-1} \left(c_k \left(\frac{d}{dz}\right)^k
 \dt (z) + d_k \frac{\rho_k (z ; \bar x; \bar y; \bar
 t)}{z^m }\right),~~~d_{m-1}\neq 0,\nonumber\\
  \tilde \rho_{\ell} &=&
  \rho_{\ell - 1} (z; \bar x, \bar y, \bar t).
 \eea
 \end{corollary}
 Before proving theorems 5.1 and 5.2 and corollary 5.3, we need a
 crucial Lemma :

 %\newpage

 \begin{lemma}  The following two identities hold for
 the $m$-periodic sequences of weights of (5.1):

\bigbreak

 \noindent $\displaystyle
 \BX_1 (\lb) D_n (\rho)$
 \begin{eqnarray}
  &=& e^{\sum_1^{\infty} \bar t_{im} \lb^{im} }
 e^{\sum_1^{\infty} \bar x_i \lb^i } D_n ((\lb - z) \rho)
 \nonumber\\
 \noindent &=& e^{\sum_{1}^{\infty}
  \bar t_{im} \lb^{im}} e^{\sum_{1}^{\infty} \bar x_i \lb^i }
(-1)^n \det
 (\la z^i ,\dt (z - \lb) \ra, \la z^i ,\rho_0 \ra ,
 \ldots,\la z^i ,
 \rho_{n-1} \ra)_{0 \leq i \leq n},\nonumber\\
 \end{eqnarray}
 \noindent $\displaystyle \Lb^{-1}
  e^{\sum_{i=1}^{\infty} \bar t_{im} \lb^{im}}
    \BX_2 (\lb)
 D_n (\rho)$
 \begin{eqnarray}
 &=& e^{\sum_{1}^{\infty} \bar y_i \lb^i}
 (-1)^{n-1} \nonumber\\
 &&~~~~~~~~\det \left(\left\la z^i ,
 \frac{\sum_{r=0}^{m-1} \lb^{m-r} \rho_r }{\lb^m - z^m }
 \right\ra,
 \la z^i ,\rho_0 \ra, \ldots, \ldots
 \la z^i ,\rho_{n-2} \ra\right)_{0 \leq i \leq n-1},\nonumber\\
 \end{eqnarray}
with all the $\rho_j$'s in the determinants above
evaluated at $\bar x, \bar y, \bar t$ according to
formula (5.1).
 \end{lemma}

 \proof : Here we use the first solution
  $m_{\infty}$ of (3.4)
   (and its calculation in the proof of Theorem 3.1),
 and in
 the second equality, we use the familiar formula
 $e^{-\sum u^i/i } = 1 - u$.  So, one computes, using
$\BX_1(\lb)$, as defined in (4.1):

\bigbreak

 \noindent $\displaystyle \BX_{1} (\lb) D_n
 \Bigl(\rho(z;\bar x,\bar y,\bar t)\Bigr)$
 \begin{eqnarray*}
 &=& \lb^n
 e^{\sum_{\ell = 1}^{\infty} \bar t_{\ell m} \lb^{\ell m} }
 e^{\sum_{1}^{\iy} \bar x_i \lb^i }
 e^{-\sum_{1}^{\infty} \frac{\lb^{im} }{im}
 \frac{\pl}{\pl \bar t_{im}} }
 e^{- \sum \frac{\lb^{-i}}{i} \frac{\pl}{\pl \bar x_i } }\\
 \noindent && ~~~~~\cdot \det \left\{ ( \la z^i ,\rho_j
 (z;0,0,0)
 e^{\sum \bar x_r z^r }
 e^{\sum_{\ell=1}^{\infty}
 \bar t_{\ell m} z^{\ell m}} \ra )_{0 \leq i,j \leq \infty}
 e^{- \sum_{1}^{\iy} \bar y_r \Lb^{r \top }}
 \right\}_{0 \leq i,j \leq n-1}\\
 \noindent &=& \lb^n e^{\sum_{\ell=1}^{\infty}
 \bar t_{\ell m} \lb^{\ell m}} e^{\sum \bar x_i \lb^i}\\
 \noindent & &\det \left\{ \left( \left\la z^i , \rho_j
 (z,0,0,0)
 e^{-\sum_{1}^{\infty} \frac{1}{r} (\frac{z}{\lb})^r}
 e^{\sum_1^{\infty} \bar x_r z^r }
 e^{\sum_1^{\infty} \bar t_{\ell m}
 z^{\ell m} } \right\ra \right)_{0 \leq i,j \leq \infty}
 e^{- \sum_1^{\infty} \bar y_r \Lb^{r \top} }
 \right\}_{0 \leq i,j \leq n-1 } \\
 &=& e^{\sum_{\ell=1}^{\infty} \bar t_{\ell m} \lb^{\ell m}} e^{\sum
 \bar x_i \lb^i}
 D_{n} \Bigl((\lb - z) \rho(z;\bar x,\bar y,\bar t )\Bigr),
 \end{eqnarray*}

 \noindent upon bringing $\lb^n $ in the $n \times n$ determinant, and using
 again the first expression (3.4) for $m_{\infty}$.
 But, using (1.5) and (1.14), we compute, where in
 this calculation
 $\rho_i:=\rho_i(\bar x, \bar y,\bar t)$,
 \begin{eqnarray*}
 D_{n} \left((\lb - z) \rho\right)
 &=& \det ( \left\la z^i , (\lb - z) \rho_0 \right\ra ,
 \ldots, \left\la z^i , (\lb - z) \rho_{n-1} \right\ra)
 _{0 \leq i \leq n-1}
 \mbox{}\\
 &=& \det \left(\la z^i ,\rho_0 \ra, \ldots, \la z^i, \rho_{n-1}\ra ,
 \lb^i \right)_{0 \leq i \leq n} ,\mbox{ using Corollary 2.3},\\
 &=& (-1)^n \det \left(\left\la z^i , \dt (z - \lb ) \right\ra, \la z^i
,\rho_0 \ra
 ,\ldots, \la z^i ,\rho_{n-1} \ra\right)_{0 \leq i \leq n},
 \end{eqnarray*}

 \noindent using the $\dt$-function property, thus establishing
 the identity (5.13).

 For future use, we shall need the following
 easy identities :
\be
 e^{\sum_{i=1}^{\infty} \frac{a^{im}}{im} } =
 e^{\frac{1}{m} \sum_1^{\infty} \frac{(a^m)^i }{i} } =
 (\frac{1}{1 - a^m })^{1/m} ,
\ee
 and (in the exponential, one sums over $i$'s, not
 multiples of $m$)
\begin{eqnarray}
 e^{\sum_{\stackrel{m \not | i}{i=1} }^{\infty}
 \frac{a^i}{i} }
 &=& e^{\sum_1^{\infty} \frac{a^i}{i} } e^{- \sum_1^{\infty}
 \frac{a^{im}}{im} }\nonumber\\
 &=& \frac{(1 - a^m )^{1/m} }{1 - a} \nonumber\\
 &=& \frac{1 - a^m }{1 - a} (1 - a^m )^{-1 + 1/m }
 \nonumber\\
 &=& \sum_0^{m-1} a^i (1 - a^m )^{-1 + 1/m } .
 \end{eqnarray}
Notice that, for any moment matrix $m_{\iy}$ defined
by $m$-periodic weights,
  $$
\left(m_{\iy}\left(\frac{\Lb^{\top}}{\lb}\right)^n\right)_{ij}
=
 \frac{\mu_{i,j + n}}{\lb^n} =
\left\la  z^i , \frac{\rho_{j + n}}{\lb^n} \right\ra  ;
$$
in particular, using the periodicity of
the sequence $\rho_j=\rho_j(z;0,0,0)$, we
have
$$
\left(m_{\iy}\left(\frac{\Lb^{\top}}{\lb}\right)^{mk}
\right)_{ij} =
\left\la  z^i , \frac{\rho_{j+mk}}{\lb^{mk}} \right\ra   =
\left\la  z^i , \left(\frac{z}{\lb}\right)^{mk} \rho_j
\right\ra  .
$$
Combining these two facts, we find
\be
\left(m_{\iy}\left(\frac{\Lb^{\top}}{\lb}\right)^{r} f
\left(\left(
\frac{\Lb^{\top}}{\lb}\right)^m\right)\right)_{ij} =
\left\la  z^i , f
\left(\left(\frac{z}{\lb}\right)^m\right)
\frac{\rho_{j+r}}{\lb^r} \right\ra  . \ee
 Now using
$\BX_2(\lb)$, defined in (4.1) and using (3.4) for
$m_{\iy}$, compute:
\begin{eqnarray*}
 \lefteqn{  \Lb^{-1} e^{\sum^\iy_{i=1}
t_{im} \lb^{im}} \BX_2 (\lb)D_n\Bigl(\rho
 (z , \bar x , \bar y , \bar t)\Bigr)}\\ &&\\
 &=&
\lb^{1-n} e^{\sum \bar y_i \lb^i} e^{\sum^\iy_1
\frac{\lb^{-rm}}{rm} \frac{\pl}{\pl \bar t_{rm}}}
e^{-\sum^\iy_1 \frac{\lb^{-r}}{r} \frac{\pl}{\pl \bar
y_r}}
 \\ & &
 \hspace{1cm}\det \left\{  \left\la z^i , \rho_j (z;0,0,0)
 e^{\sum^\iy_{r=1}\bar t_{rm} z^{rm}}
 e^{\sum \bar x_r z^r} \right\ra_{0 \leq i,j < \iy}
 e^{- \sum \bar y_r
 \Lb^{Tr}} \right\}_{0 \leq i,j \leq
 n-1}
\\ &=&
\lb^{1-n} e^{\sum \bar y_i \lb^i}
 \det \Biggr\{ \left\la z^i, \rho_j (z;0,0,0)
 e^{\sum^{\iy}_{r=1}
 \frac{1}{\ell m}\left(\frac{z}{\lb}\right)^{\ell m}}
 e^{\sum^{\iy}_{r=1} \bar t_{rm} z^{rm}}
 e^{\sum \bar x_r z^r}
 \right\ra_{0 \leq i,j < \iy}
\\ & &
 \hspace{5cm} e^{\sum_{m \not | r}\frac{1}{r}
 \left(\frac{\Lb^{\top}}{\lb}\right)^r}
 e^{-\sum \bar y_r
 \Lb^{\top r}} \Biggr\}_{0 \leq i,j < n-1}
\\ &=&
 \lb^{1-n} e^{\sum \bar y_i \lb^i}
  \\
 & &
 \\
 & &
 \det\Biggl\{  \left\la z^i ~,~
e^{\sum^{\iy}_1\bar t_{rm} z^{rm}} e^{\sum \bar x_r
z^r} \frac{\rho_j(z ;0,0,0)}{(1-(z/\lb)^m)^{1/m}}
 \right\ra_{0 \leq i,j < \iy}
 \\  & & \hspace{1cm}
\frac{\sum^{m-1}_0 \left(
\frac{\Lb^{\top}}{\lb}\right)^i}{\left( 1-\left(
\frac{\Lb^{\top}}{\lb}\right)^m\right)^{1-1/m}}
e^{-\sum^{\iy}_1 \bar y_r \Lb^{\top r}} \Biggr\}_{0
\leq i,j \leq n-1} , ~ \mbox{using}~ (5.15)
~\mbox{and}~(5.16),
 \\  &=&
\lb^{1-n} e^{\sum \bar y_i \lb^i}
 \\
 &&\det \Biggl\{  \left\la z^i ,
 e^{\sum^{\iy}_1 \bar t_{rm} z^{rm}}
 e^{\sum \bar x_r z^r}
 \frac{\sum^{m-1}_{r=0} \frac{\rho_{j +
 r}(z;0,0,0)}{\lb^r}}
 {\left(1-\left(\frac{z}{\lb}\right)^m\right)^{1/m}
 \left(1-\left(\frac{z}{\lb}\right)^m\right)^{1-1/m}}
  \right\ra_{0 \leq  i,j < \iy}
 \\ & & \hspace{6cm}
e^{-\sum \bar y_r \Lb^{\top r}} \Biggr\} _{0 \leq i,j
\leq n-1} ,~ \mbox{using}~ (5.17),
\\ &=&
\lb^{1-n} e^{\sum \bar y_i \lb^i}
\det \Biggl\{  \left\la z^i ,
e^{\sum^\iy_1 \bar t_{rm} z^{rm}} e^{\sum
\bar x_r z^r}
\frac{\sum^{m-1}_{r=0} \lb^{m-r} \rho_{j+r} (z;0,0,0)}
 {\lb^m - z^m} \right\ra_{0 \leq i,j < \iy}
\\ & & \hspace{6cm}
 e^{-\sum \bar y_r \Lb^{\top r}} \Biggr\}_{0 \leq i,j \leq n-1}
 \\ &=&
 \lb e^{\sum \bar y_i \lb^i} \det \left\{ \left\la z^i ,
 \sum^{m-1}_{r=0}
 \frac{\lb^{m-1-r}}{\lb^m - z^m}
 \rho_{j+r} (z , \bar x , \bar y , \bar t)
 \right\ra_{0 \leq i,j \leq n-1}\right\}
 \\ &=&
 \lb e^{\sum \bar y_i \lb^i} (-1)^{n-1}
 \det \Biggl( \left\la z^i , \sum^{m-1}_{r=0}
  \frac{\lb^{m-1-r} \rho_r
 (z;\bar x,\bar y,\bar t)}{\lb^m - z^m} \right\ra , \left\la z^i ,
 \rho_0 (z;\bar x,\bar y,\bar t)\right\ra ,
\\ & & \hspace{6cm}
\cdots, \la z^i , \rho_{n-2} (z;\bar x,\bar y,\bar
t)\ra \Biggr)_{0 \leq i \leq n-1} .
\end{eqnarray*}
The second from the last expression is a consequence
of (3.4) and (3.5), according to the argument in the
proof of Theorem 3.1 and the linearity of (3.5) with
respect to the measures $\rho=(\rho_0,\rho_1,...)$,
while the last line is obtained by replacing the jth
column $C_j$ by $C_j
-
\lb C_{j-1} , 2 \leq j \leq n$, in the previous
determinant and using the identity:
\begin{eqnarray*}
%\hspace{2cm}
 \sum^{m-1}_{r=0}
 \frac{\lb^{m-1-r} \rho_{j+r}}{\lb^m - z^m}
  - \lb
 \sum^{m-1}_{r=0} \frac{\lb^{m-1-r}
 \rho_{j+r-1}}{\lb^m - z^m}
 &=& \frac{\rho_{j+m-1} - \lb^m \rho_{j-1}}{\lb^m - z^m}
 \\ \hspace{2cm} &=&
\frac{z^m \rho_{j-1} - \lb^m \rho_{j-1}}{\lb^m - z^m}
 \\ \hspace{2cm} &=&
  - \rho_{j-1}
\end{eqnarray*}
\qed

\underline{{\sl Proof of Theorem 5.1}} From theorem
4.1 (the map (4.4)), and from (5.13) of Lemma 5.4, it
follows that
 \bean
 \tau_n=D_n(\rho)\mapsto
  \tilde \tau_n &=& \sum^{m-1}_{k=0}
 a_k \BX_1 (\om^k \lb) D_n(\rho)
\\
&=& \sum^{m-1}_{k=0} e^{\sum^\iy_{1=i} \bar t_{im}
 \lb^{im}}
 e^{\sum^\iy_{i=1} \bar x_i (\om^k \lb)^i} a_k
 D_n ((\om^k \lb - z)\rho)
 \\
  &=& \sum^{m-1}_{k=0} \tilde a_k ~D_n
  \left((\om^k \lb - z)\rho\right)
\\
&=& (-1)^n \sum^{m-1}_{k=0} \tilde a_k \det \left( \la
z^i , \dt (z - \om^k \lb)\ra , \la z^i , \rho_0 \ra ,
\cdots , \la z^i , \rho_{n-1}\ra \right)_{0 \leq i
\leq n} . \eean The expression on the right hand side
of the third identity establishes the second identity
(5.6), whereas the last identity establishes the third
(5.6), ending the proof of Case 1. Setting all but one
$a_k=0$, establishes (5.4) in Case 1.  \qed

\underline{\sl Proof of Theorem 5.2} : According to
Theorem 4.2 and Lemma 5.4, the UL-Darboux transform
(5.8) with $\beta_n$, given in (5.9) acts on $\tau_n
 (z ; \bar x , \bar y , \bar t):
 = D_n(\rho(z ; \bar x , \bar y , \bar t))$ as
 follows:
\bean \tau_n \longmapsto & \tilde \tau_n &\\ &=&
\left( \Lb^{-1} \tau \Phi (\lb)\right)_n
 \\ &=&
  \left(
  \sum^{m-1}_{k=0} \left( a_k \Lb^{-1} \BX_1 (\omega^k \lb) + b_k
 \Lb^{-1} e^{\sum^\iy_{i=0} \bar t_{im}
 \lb^{im}} \BX_2 (\om^k \lb)\right) \tau \right)_n
 \\
 &=&
  (-1)^{n-1} \det \Biggl( \la z^i ,
  \sum^{m-1}_{k=0} \tilde a_k
 \delta (z-\om^k \lb)\ra ,
\la z^i , \rho_0
 \ra , \cdots , \la z^i ,
 \rho_{n-2} \ra \Biggr)_{0 \leq i \leq n-1} \\
& & + (-1)^{n-1} \det \Biggl( \la z^i ,
 \sum^{m-1}_{k=0} b'_k
 \sum^{m-1}_{r=0} \frac{(\om^k \lb)^{m-r}}{\lb^m -
 z^m} \rho_r \ra ,
  \\
& &\hspace{6cm}
 \la z^i ,
  \rho_0 \ra , \cdots, \la z^i , \rho_{n-2} \ra
 \Biggr)_{0 \leq i \leq n-1}
 \\
& &\mbox{with} \ \tilde a_k \ \mbox{as in (5.11) and}
\ b'_k = b_k e^{\sum^\iy_1 \bar y_i (\om^k \lb)^i} ,\\
&=& (-1)^{n-1} \det \Biggl( \left\la z^i ,
\sum^{m-1}_{k=0} \tilde a_k \dt (z-\om^k \lb) +
\sum^{m-1}_{r=0} \frac{\lb^{m-r}}{\lb^m - z^m} \left(
\sum^{m-1}_{k=0} b'_k \om^{-kr}\right) \rho_r
\right\ra ,
\\
&&\hspace{7cm}
 \la z^i ,
\rho_0 \ra , \cdots, \la z^i , \rho_{n-2} \ra \Biggr)_{0 \leq i \leq n-1}
\\
&=& (-1)^{n-1} \det \left( \la z^i , \tilde \rho_0 \ra
, \la z^i , \tilde \rho_1 \ra , \cdots , \la z^i ,
\tilde \rho_{n-1} \ra \right)_{0 \leq i \leq n-1} ,
\eean using the new $\tilde \rho_i$ defined in (5.10).

 Finally, %setting $(\bar x, \bar y, \bar t)=(0,0,0)$,
 using the $\dt$-function property in the second
 identity, and using $\tilde\rho_k=\rho_{k-1}$
 for $k$ not a multiple of $m$% \not | k$
 , we prove
 the new sequence is generalized $m$-periodic:
\bean
z^m \tilde \rho_0 &=&
  \sum^{m-1}_{k=0} \left(
 \tilde a_k z^m \dt (z-\om^k
 \lb) + \tilde b_k
  \frac{\lb^m +
 (z^m-\lb^m )}{z^m-\lb^m } \rho_k (z)\right)
\\ &=&
 \lb^m
 \sum^{m-1}_{k=0}   \left( \tilde a_k \dt
 (z-\om^k \lb) + \tilde b_k
 \frac{\rho_k (z)}{z^m-\lb^m } \right)
 %\\ & &\hspace{7cm}
 +\sum^{m-1}_{k=0} \tilde b_{k} \rho_k (z)
\\
                  &=& \lb^m \tilde \rho_0(z) +
                  \sum^m_{k=1}
\tilde b_{k-1}  \tilde \rho_k (z)
\\
& \in & \
 \mbox{span} \  \{ \tilde \rho_0 ,
 \cdots, \tilde \rho_m \},~~\mbox{with the condition that}~
 \tilde b_{m-1}\neq 0,\\
& & \\ z^m \tilde \rho_k &=& z^m \rho_{k-1}
 =\rho_{k-1+m} =\tilde \rho_{k+m},~~\mbox{for}~k\geq 1,
 ~\mbox{not a multiple of $m$},
\eean establishing Theorem 5.2.\qed

\remark  As already pointed out in the remark
following the statement of Theorem 5.2, although the
sequence $\rho(\bar x, \bar y, \bar t)$ is generalized
$m$-periodic in the sense of definition 3.2, it is not
$m$-periodic in the sense of (0.1) and it only leads
to a solution $m_{\iy}$ of (3.2) in the $\bar x$ and
$\bar t$ variables, but not in $\bar y$. However,
since the matrix $\tilde L$ is computed from the new
polynomials $p_n(z;\bar x, \bar y,\bar t)$ (defined in
Theorem 3.1), by $z^mp=\tilde L p$ and since
establishing the form of $p_n$ only depended on the
$x$-dependence of $\tau$ through $\rho(\bar x, \bar y,
\bar t)$, it is indeed defined by $m$-periodic
weights.

\bigbreak

\underline{\sl Proof of Corollary 5.3} : The proof
follows at once from theorem 5.2 by letting $\lb \to
0$, and $b_k \to \iy$, and by picking appropriate
$a_k$. \qed

\underline{\sl Proof of Theorem 0.1, 0.2 and Corollary
0.3 } :
 The proofs follow from setting
  $(\bar x, \bar y,\bar t)=(0,0,0)$ in Theorems 5.1, 5.2, and Corollary
 5.3.

\newpage

\section{Example 1: Darboux transform for tridiagonal
matrices }

%$\tinydot$

In this section, we specialize to the case $m=1$,
which
        leads naturally to orthogonal polynomials,
        to three-step relations,
and so to semi-infinite tridiagonal matrices $L$. The
LU-Darboux transform on such matrices consists of
decomposing the matrices $L-\lb I$ as a product of
lower- and upper-triangular matrices and multiplying
them in the opposite order. The UL-Darboux goes the
other way around. Unlike the case of bi-infinite
matrices, the LU-Darboux map for the semi-infinite
case is a unique operation, of course depending on the
parameter $\lb$, whereas UL-Darboux depends on a free
parameter $\sigma$, besides $\lb$.

 What is the effect of this operation on weights? Theorems
 5.1 and 5.2 show that
LU-Darboux has the effect of multiplying the weight
$\rho(z)$ with $\lb -z$ and UL-Darboux divides the
weight by $\lb -z$, augmented by a delta-function
$(\sigma/\lb)\delta(z-\lb)$ involving the free
parameter $\sigma$.

In \cite{AvM5}, we have shown that, upon letting the
tridiagonal, bi-infinite matrices flow according to
the standard Toda lattice, the LU- or UL-Darboux
transforms act on the eigenvectors as discrete
Wronskians and on the $\tau$-functions as vertex
operators especially taylored to the Toda lattice.
Both transforms depend on one free (projective)
parameter. The reduction to the semi-infinite case
cuts out this freedom for the LU-transform, but not
for the UL-transform.

This vertex operators technology can be used very
efficiently to get the results, after setting $t=0$;
in fact one can establish a dictionary between the
three points of view: {\em weights, vertex operators
and Darboux transforms}, as summarized in (0.23); the
point of the dictionary is contained in the subsequent
theorems and corollaries. The relationship rests on an
elementary {\em addition formula}; namely, the sum of
moment determinants $D_n$ and $D_{n-1}$ with regard to
specific weights is again a moment determinant $D_n$,
but with respect to a new weight: $$ D_n(\rho)+c
D_{n-1}\left( (\lb-z)^2 \rho(z)\right)= D_n(\rho(z)+c
\delta(\lb-z)); $$ this fact is not surprising, in
view of the fact that if the $\tau=(\tau_n)_{n\geq 0}$
is a vector of $\tau$-functions for the standard Toda
lattice, then the following expressions $$
\tau(t)+c\BX(t,\lb)\tau(t) $$ forms a Toda
$\tau$-vector as well, where $\BX(t,\lb)$ is the
standard Toda vertex operator, defined in (0.21), and
acting on $\tau$ as in (0.22).

An arbitrary weight $\rho(z)$ on $\BR$ yields a
$1$-periodic sequence
$(\rho(z),z\rho(z),z^2\rho(z),\ldots)$ and a moment
matrix $m_{\iy}$, satisfying $\Lb
m_{\iy}=m_{\iy}\Lb^{\top}$ (H\"ankel matrix). Also
%
% its
%determinant\footnote{$D_0=1$}
%$D_n(\rho)$
\be
m_n(\rho)=(\mu_{i+j}(\rho))_{0\leq i,j\leq
n-1},~~D_n(\rho)=\det m_n(\rho),
\mbox{\,\,with\,\,}\mu_k(\rho)=\int_{\BR}z^k\rho(z)dz,
\ee
 with $D_0=1$. The orthogonality relations (3.7)
 lead to monic orthogonal polynomials in $z$ of
 degree $n$
 %, i.e., $\la p_i,p_j \rho
 %\ra=\delta_{ij}h_i$, with
\be
p_n(z)=\frac{1}{D_n(\rho)}\det\left(
\begin{tabular}{lll|l}
$\mu_0(\rho)$&...&$\mu_{n-1}(\rho)$&1\\ $\vdots$&
&$\vdots$&$\vdots$\\
$\mu_{n-1}(\rho)$&...&$\mu_{2n-2}(\rho)$&$z^{n-1}$\\
\hline $\mu_n(\rho)$&...&$\mu_{2n-1}(\rho)$&$z^n$
\end{tabular}\right),~~\mbox{with}~~
 \la p_i,p_j \rho
 \ra=\delta_{ij}h_i
\ee In turn, the semi-infinite vector of polynomials
$p=(p_n(z))_{n\geq 0}$ leads to a semi-infinite
tridiagonal matrix $L$, defined by
\be
zp=Lp,\mbox{\,\,with\,\,}L=\left(\begin{tabular}{lll}
$b_0$&1& \\ $a_0$&$b_1$&$\ddots$\\
 &$\ddots$&$\ddots$
\end{tabular}
\right). \ee

\begin{theorem}
 (i) Given the weight $\rho(z)$ and $\lb \in \BC$, the
eigenvector of $L$, corresponding to the eigenvalue
$\lb$,
\be
\left(\Phi_n(\lb)\right)_{n\geq 0}=\left( p_n(\lb)
\right)_{n\geq 0}=
\left(\frac{D_n((\lb-z)\rho(z))}{D_n(\rho)}\right)_{n\geq
0} \in (L-\lb I)^{-1}(0,0,0,...)\ee specifies a unique
%{\em lower-upper}
 {\em LU-Borel factorization} $$
L-\lb I=L_- L_+= \left(\begin{array}{ccc} 1 &0&\\
\al_0&1&\ddots\\ &\ddots&\ddots
\end{array}\right)
\left(\begin{array}{ccc} \beta_{0} &1&\\
0&\beta_1&\ddots\\
 &\ddots&\ddots
\end{array}\right),
$$ with
\be
\beta_n:= -\frac{\Phi_{n+1}(\lb)}{\Phi_{n}(\lb)},~~~~
\al_{n-1}=b_n-\beta_n-\lb. \ee
 The {\em LU-Darboux transform}
\be
L-\lb=L_-L_+\longmapsto \tilde L-\lb=L_+L_-, \ee
induces the following map on weights $\rho(z)$:
\be
\rho(z)  \longmapsto \rho(z)(\lb -z) \ee (ii) The
two-dimensional eigenspace, corresponding to the
eigenvalue $\lb$ and with a different boundary
condition at $n=0$ , is given by
%with parameter $\sigma$
\be
\left(\Phi_n(\lb)\right)_{n\geq 0}
=\left(\frac{\frac{\sigma}{\lb}D_n((\lb-z)\rho(z))+
D_{n+1}\left(\frac{\rho(z)}{\lb-z}
\right)}{D_n(\rho)}\right)_{n\geq 0}\in (L-\lb
I)^{-1}(1,0,0,...). \ee It specifies a
$\sigma$-dependent family of
%{\em upper-lower}
 {\em UL-Borel factorizations},
\be
L-\lb=L'_+ L'_- =~\left(\begin{array}{ccc} \al_{-1}
&1&\\ 0&\al_{0}&\ddots\\ &\ddots&\ddots
\end{array}\right)~\left(\begin{array}{ccc}
1 &0&\\ \beta_{0}&1&\ddots\\ &\ddots&\ddots
\end{array}\right),
\ee with the same $\beta_n$ and $\al_{n-1}$ as in
(6.5), but with $\Phi_n$ defined by (6.8). This
defines {\em UL-Darboux transforms}
\be
L-\lb=L'_+L'_-\longmapsto \tilde L'-\lb=L'_-L'_+, \ee
inducing the following map on weights $\rho(z)$:
\be
\rho(z) \longmapsto \left(\frac{\rho(z)}{\lb
-z}+\frac{\sigma}{\lb} \dt(\lb-z)\right), \ee
\end{theorem}

\proof These statements follow immediately from
setting $m=1$ in Theorems 0.1 and 0.2.

\vspace{0.3cm}

\begin{corollary} Consider the map $L\longmapsto
L''$, defined by a UL-Darboux transform followed by a
LU-transform: $$ L-\lb=L_+ L_- \longmapsto L'-\lb:=L_-
L_+\longmapsto L'-\mu=L'_- L'_+\longmapsto
L''-\mu:=L'_+ L'_-, $$ where the parameter of the
first UL-Darboux map is given by $$ \sigma := \frac{c
\mu}{\mu-\lb} ; $$ then, upon taking the limit $\mu
\longrightarrow \lb$, the map above induces a map of
weights $$ \rho(z) \longmapsto \rho(z)+c \dt(\lb-z).
$$
\end{corollary}

\vspace{0.3cm}

\begin{corollary}
Concatenating $m$ LU-Darboux transforms with parameter
$\mu_i$ and $n$ UL-Darboux transforms with $n_i$
parameters converging to $\lb_i$ ($n_1+...+n_r=n$),
induces a map of weights: $$ \rho(z) \longmapsto
\left(\frac{\prod_1^m(z-\mu_i)}{\prod^r_1(z-\lb_k
)^{n_k}}\rho(z) +\sum_{k=1}^r \sum_{j=1}^{n_k}
c_{kj}\left(\frac{\pl}{\pl z}\right)^{j-1}
\dt(z-\lb_k)\right). $$ Upon picking the $\mu_i$
appropriately, the fraction in front of $\rho(z)$ in
the formula above disappears.
\end{corollary}

These statements are established by letting the moment
matrix $m_{\iy}$ flow according to (1.1), and then
letting the associated tridiagonal matrix $L$ flow
according to the standard Toda lattice (remember
$(L^n)_+$ denotes the strictly upper-triangular part
of $L^n$)
\be
\frac{\pl L}{\pl t_n}=[(L^n)_+,L],\quad n=1,2,...~.
\ee In the 3-reduction of 2-Toda, only one set of
times $t=\bar t=(t_1,t_2,\ldots)$ of (2.1) remain. The
$(\bar x, \bar y,\bar t)$ evolution (3.5) of the
weight $\rho(z)$ reduces to the simple formula
$$\rho_t(z):=e^{\sum^{\iy}_1 t_i z^i}\rho(z),$$ which
was shown in a direct way in \cite{AvM2}, for
instance;in other terms, the Toda vector fields (6.12)
linearize at the level of the weight $\rho_t(z)$.
 The
deformations $\rho_t(z)$ of $\rho(z)$ enable one to
define $t$-dependent moments $\mu_k(\rho_t(z))$,
associated moment matrices $m_n(\rho_t(z))$, and
$t$-dependent monic orthogonal polynomials $p_n(z;t)$
of degree $n$, with $L^2$-norms
\be
h_n(t):=\int_{\BR} p_n^2(t,z)\rho(t,z)
dz=\frac{\tau_{n+1}(t)}{\tau_n(t)}. \ee
 The entries of the $t$-dependent $L$-matrix are expressed in terms
 of the $\tau$-functions
\be
D_n(\rho_t)=\det m_n(\rho_t)=:\tau_n(t), \ee
 as follows,
\be
b_k=\frac{\pl}{\pl
t_1}\log\frac{\tau_{k+1}}{\tau_k}\quad\mbox{and}\quad
a_{k-1}= \frac{\tau_{k-1}\tau_{k+1}}{\tau^2_k}. \ee
 %in
%terms of the sequence of $\tau$-functions $\tau_n$,
%themselves related by bilinear identities to be
%written later in (1.5). For later use, also define the

Setting $m=1$ in the {\em vertex operators}
$\BX_1(t,\lb)$ and $\BX_2(t, \lb)$ of (4.1) leads to
\be
\BX_1(t,\lb):=\chi(\lb)X(t,\lb)\quad\mbox{and}\quad\BX_2(t,\lb)
:=\chi(\lb^{-1}) X(-t,\lb)\Lambda.
 \ee
They are generating functions of symmetries of the
standard Toda Lattice and act on $\tau$-vectors; see
\cite{AvM5}. The vertex operator
 $\BX(t,\lb)$, defined in (0.21), is obtained
from $\BX_1(t,\lb)$ and $\BX_2(t,\lb)$, as follows
  \be
\BX(t,\lb) :=\lim_{\mu \rightarrow \lb}
\frac{1}{1-\lb/\mu}\left(e^{\sum t_i
\mu^i}\BX_2(t,\mu)\right)^{-1}\BX_1(t,
\lb)=\Lb^{-1}\chi(\lb^2)e^{\sum
t_i\lb^i}e^{-2\sum\frac{\lb^{-i}}{i} \frac{\pl}{\pl
t_i}}%\nonumber\\ &=&\BX(t,\lb)
\ee
 has the surprising property (in view of the
 non-linearity of the problem) that,
  given a vector $\tau=
 (\tau_0,\tau_1,\ldots)$ of Toda
 $\tau$-functions, the new vector (see (0.22))
 \be\tau + \BX(t,\lb)\tau
  \ee
 is a new vector of Toda
 $\tau$-functions. For connections with vertex operator algebras, see V.
Kac \cite{Kac}.

The following statements, Theorem 6.4 and Corollary
6.5 are completely parallel with Theorem 6.1 and
Corollary 6.2. They provide a {\em dictionary},
between the three points of view:

\begin{theorem}  (i) The eigenvector\footnote{with
asymptotics $\Phi_n(t,\lb)= e^{\sum t_i \lb^i} \lb^n
(1+O(\lb^{-1}))$.} \bea \Phi(t,\lb):=
\frac{\BX_1(t,\lb)\tau(t)}{\tau(t)}%\nonumber\\
 &=&
e^{\sum^{\iy}_{0} t_i \lb^i}
\left(\frac{D_n((\lb-z)\rho_t(z))}
{D_n(\rho_t)}\right)_{n\geq 0}\nonumber\\ &  \in &
(L(t)-\lb I)^{-1}(0,0,0,...) \eea induces a LU-Borel
factorization, as in (6.5), with
 $$
\al_n=\frac{\pl}{\pl t_1}\log\Phi_{n+1}(t,\lb)-\lb $$
and
\be
\beta_n=- \frac{\Phi_{n+1}(t,\lb)}{\Phi_n(t,\lb)}=
-\frac{\pl}{\pl t_1} \log \left(\frac{\tau_n
}{\tau_{n+1}}\Phi_n(t,\lb)\right); \ee the LU-Darboux
transform $L(t)-\lb \mapsto \tilde L(t)-\lb$ with new
entries $\tilde b_n$ and $\tilde a_n$, is given by
(6.6) in terms of the new $\tau$-function:
\be
\tau\longmapsto\tilde \tau=\tau\Phi=\BX_1(t,\lb)
\tau(t). \ee (ii) The eigenvectors \bea
\Phi(t,\lb):&=& \frac{1}{\lb}
\frac{\left(\sigma\,\BX_1(t,\lb)+~e^{\sum
t_i\lb^i}\BX_2(t,\lb)\right)\tau(t)}{\tau(t)}
\nonumber\\ &=&\left(\frac{\frac{\sigma}{\lb}e^{\sum
t_i\lb^i}
D_n((\lb-z)\rho_t(z))+D_{n+1}\left(\frac{\rho_t(z)}{\lb-z}
\right)}{D_n(\rho_t)}\right)_{n\geq 0}\nonumber\\ &\in
& ~(L-\lb I)^{-1}(1,0,0,...) \eea induce a
UL-factorization with $\al$ and $\beta$ as in (6.20),
but with $\Phi_n(t,z)$ defined in (6.22); it defines a
UL-Darboux transform $L(t)-\lb \mapsto \tilde
L'(t)-\lb$, as in (6.10), with new entries $\tilde
b'_n$ and $\tilde a'_n$, given by (6.15) in terms of
the new $\tau$-function
\be
\tau\longmapsto\tilde
\tau'=\Lb^{-1}\lb\tau\Phi=\Lambda^{-1}\left(\sigma~\BX_1(t,\lb)+\,e^{\sum
t_i\lb^i} \BX_2(t,\lb)\right)\tau(t). \ee
\end{theorem}

\begin{corollary} Consider the map $L(t)\longmapsto
L''(t)$, defined by a UL-Darboux transform followed by
a LU-transform, as in Corollary 6.2, with that same
choice of $\sigma$. It induces the map (6.18) at the
level of Toda $\tau$-vectors:
\be
D_n(\rho_t)  \longmapsto  D_n\left(\rho_t(z)
+ce^{\sum_1^{\iy}t_iz^i} \dt(\lb-z)\right)
=\left(1+c\BX(t,\lb)\right)D_n(\rho_t), \ee where
$\BX(t,\lb)$ is the Toda lattice vertex operator
(6.17).
\end{corollary}

Instead of using Theorems 0.1 and 0.2 to establish
those results, one can prove them directly, using the
formulae in proposition 6.6 below. In this way,
classical formulae have a natural $\tau$-function
counterpart.

%\newpage

\begin{proposition} Given the weights $\rho_t(z)$, the
moments $\mu_i(\rho_t(z))$ and the $\tau$-functions
$\tau_n(t):=D_n(\rho_t)$, we have the following
expressions for\footnote{Remember
$[\al]:=(\al,\al^2/2,\al^3/3,...)$.}:

\noindent
\begin{itemize}
  \item \underline{the monic orthogonal
polynomials}% $p_n(u;t)$
: \bea
\hspace{-1cm}p_n(u;t)=\frac{1}{D_n(\rho_t)}\det\left(
\begin{tabular}{lll|l}
$\mu_0$&...&$\mu_{n-1}$&1\\ $\vdots$&
&$\vdots$&$\vdots$\\
$\mu_{n-1}$&...&$\mu_{2n-2}$&$u^{n-1}$\\ \hline
$\mu_n$&...&$\mu_{2n-1}$&$u^n$
\end{tabular}\right)\nonumber%\nonumber\\
&=&\frac{D_n((u-z)\rho_t(z))}{D_n(\rho_t(z))}\nonumber\\
&=&u^n\frac{\tau_n(t-[u^{-1}])}{\tau_n(t)} ;\nonumber
\eea

\bea
 \hspace{-2cm} q_{n-1}(u;t):=\int_{\BR^n}\frac{p_n(x;t)}{u-x}
\rho_t(x)dx
 &=& \frac{1}{D_{n-1}(\rho_t(z))}
 D_{n}\left( \frac{\rho_t(z)}{u-z}  \right)\nonumber\\
  &=&u^{-n}
\frac{\tau_{n}(t+[u^{-1}])}{\tau_{n-1}(t)}.\nonumber%\nonumber\\
\eea

  \item \underline{The Christoffel-Darboux
kernels}: (for $h_i$, see (6.13)) \bea \sum_{0\leq
j\leq
n}h_j^{-1}(t)p_j(u;t)p_j(v;t)&=&-\frac{1}{D_{n+1}(\rho_t)}\det\left(
\begin{array}{lllll}
0&1&v&...&v^n\\ 1&\mu_0&\mu_1&...&\mu_n\\
u&\mu_1&\mu_2&...&\mu_{n+1}\\ \vdots& & & & \\
u^n&\mu_n&\mu_{n+1}&...&\mu_{2n}
\end{array}
\right)  \nonumber\\ &=&
\frac{D_n((u-z)(v-z)\rho_t(z))}{D_{n+1}(\rho_t)}\nonumber\\
&=&(u v)^n
\frac{\tau_n(t-[u^{-1}]-[v^{-1}],\rho)}{\tau_{n+1}(t,\rho)},
\nonumber\eea
  \item \underline{The
addition formula}: \bea D_n(\rho_t(z) +c \dt(u-z))&=&
D_n(\rho_t)+c e^{\sum t_i u^i}D_{n-1}\left(
(u-z)^2\rho_t(z)\right)\nonumber\\
&=&\left(1+c\BX(t,u)\right)D_n(\rho_t). \nonumber\eea
\end{itemize}

\end{proposition}

%\bigbreak

This last identity hinges on the addition formula: For
a $n\times n$ moment matrix $m_n$, the following
identity holds: $$ \det \left(m_n(\rho) +c
\chi_n(u)\otimes \chi_n(u)\right)=\det m_n(\rho)+c
\det m_{n-1}\left((z-u)^2 \rho(z) \right), $$
 where
 $$
\chi_n(u) \otimes \chi_n(v):=\left( u^i v^j \right)_{0
\leq i,j \leq n}. $$

%\newpage

\section{Example 2: ``Classical" polynomials, satisfying
$2m+1$-step relations}

A very natural set of ``{\em classical}" examples is
to start from a weight for the standard orthogonal
polynomials, thus corresponding to a tridiagonal
matrix $\LR_1=\LR_2$. Then we perform two consecutive
Darboux transforms on the $2m+1$-diagonal matrix
$L=\LR_1^m=\LR_2^m$. This has the effect of mapping a
$1$-periodic sequence of weights to a generalized
$m$-periodic sequence of weights, thus leading to
$2m+1$-band matrices. Therefore, one is lead to a
sequence of $2m+1$-step polynomials $\tilde p_n^{(1)}$
derived from the ``{\em standard}" ones; they satisfy
$2m+1$-step relations, i.e., $z^m \tilde p_n^{(1)}=L
\tilde p_n^{(1)}$, with $2m+1$-diagonal $L$, but not
$3$-step relations.

For a general $m$-periodic weight sequence, for
appropriate choices of $\beta$ and $\tilde \beta$, and
setting $\lb=0$ in (5.2) and (5.8), the compound map
\be L\longmapsto \tilde L=(\beta\Lb^0+\Lb)L
(\beta\Lb^0+\Lb)^{-1} \longmapsto \tilde {\tilde
L}=(\Lb^{\top}\tilde\beta+I)\tilde L
(\Lb^{\top}\tilde\beta+I)^{-1} \ee induces, according
to theorems 0.1, 0.2 and corollary 0.3, the following
compound map of weights (assuming $d_{m-1} \neq 0$):
$$ \rho \longmapsto
\tilde\rho=(z\rho_0,z\rho_1,z\rho_2,...) \longmapsto
\tilde{\tilde\rho}= \left( \sum_0^{m-1}\left( c_k
\dt^{(k)}(z)+d_k \frac{\rho_k(z)}{z^{m-1}}
\right),z\rho_0,z\rho_1,... \right). $$

A particularly interesting case is to start with
weights having the form $\rho_k(z)=z^k\rho_0(z)$,
where $\rho_0(z)$ is subjected to the following
condition: $$ \int_{\BR}\left|z^{j} \rho_0(z)\right|
dz< \iy, ~~j \geq -m+1 .$$ Then the polynomials
$p_n^{(1)}$ are orthogonal with respect to the weight
$\rho_0(z)$ and the map above becomes

\bea \rho=(z^i\rho_{0}(z))_{0\leq i<\iy} \mapsto
\tilde{\tilde{\rho}}&=&(\tilde{\tilde{\rho}}_{0},
\tilde{\tilde{\rho}}_{1}, \tilde{\tilde{\rho}}_{2},
\ldots)\nonumber\\ &=& \left( \sum_{k=0}^{m-1} \left(
c_{k}\dt^{(k)}(z)+\rho_{0}(z)\frac{d_{m-k-1}}{ z^{k}}
\right), z\rho_{0}, z^{2}\rho_{0},\ldots\right).
\nonumber\\ \eea From the general theory, this new
sequence is {\em generalized $m$-periodic} with
minimal period $m$. One checks by hand, using $z^m
\dt^{(k)}(z)=0$ for $0\leq k \leq m-1$, that \bean z^m
\tilde{\tilde{\rho}}_{0}&=&\sum_{{k=0}}^{m-1}
\left(c_{k} z^m \dt
^{(k)}(z)+d_{m-k-1}z^{m-k}\rho_{0}(z)\right) \\ &
=&\sum_{k=0}^{m-1} d_{m-k-1} z^{m-k}\rho_{0}(z)\\
 &=&\sum_{1}^m
d_{j-1}\tilde{\tilde{\rho}}_{j}.
\eean
The new moments $\tilde{\tilde{\mu}}_{ij}=\la z^i,
\tilde{\tilde{\rho}}_{j}(z) \ra $ become:
\bea
\tilde {\tilde \mu}_{i0}&=&\la z^i, \tilde {\tilde \rho}_0 \ra
 = \sum_{k=0}^{m-1} \mu_{i-k} d_{m-k-1}+
 \sum_{k=0}^{m-1} (-1)^k k! c_k \dt_{ik}\nonumber\\
 \tilde {\tilde \mu}_{ij}&=&\la z^i,
 \tilde {\tilde \rho}_j \ra=\la z^i,z^j\rho_0\ra=\mu_{i+j}~~\mbox{for} ~~j
\geq 1
,
\eea
thus defining monic polynomials $\tilde {\tilde p}_n^{(1)}(z)
 $,

\medbreak

\noindent$\displaystyle{(\det \tilde {\tilde m}_n)~
 \tilde {\tilde p}_n^{(1)}(z)}=$
\hfill
$$
\det\pmatrix{\displaystyle{\sum_{k=0}^{m-1}}\mu_{-k}d_{m-k-1}+
 \,c_0&\mu_1&\mu_2&...&1\cr
 \displaystyle{\sum_{k=0}^{m-1}}\mu_{1-k}d_{m-k-1}
 -c_1&\mu_2&\mu_3&
 ...&z\cr
 \displaystyle{\sum_{k=0}^{m-1}}\mu_{2-k}d_{m-k-1}
 +2!c_2&%
 \mu_3&\mu_4&...&z^{2}\cr %
 \vdots&\vdots&\vdots&...&\vdots \cr
 \displaystyle{\sum_{k=0}^{m-1}}\mu_{m-k-1}d_{m-k-1}
 +(-1)^{m-1}(m-1)!c_{m-1}&%
 \mu_{m}&\mu_{m+1}&...&z^{m-1}\cr
 \displaystyle{\sum_{k=0}^{m-1}}\mu_{m-k}d_{m-k-1}
 &%
 \mu_{m+1}&\mu_{m+2}&...&z^{m}\cr
 \vdots%
 &\vdots&\vdots
 &...&\vdots\cr
 \displaystyle{\sum_{k=0}^{m-1}}\mu_{n-k}d_{m-k-1}
 &%
 \mu_{n+1}&\mu_{n+2}&...&z^{n}}, $$
 which satisfy $2m+1$-step relations
 $$
 z^m p^{(1)}(z)= L p^{(1)}(z),~~\mbox{with a $2m+1$-
 band matrix $L$}.
 $$
Because of the fact that very special cases of these
 polynomials have appeared in recent work \cite{GH} on
 pentadiagonal matrices, obtained by taking squares of
 the classical tridiagonal matrices for the Laguerre and
 Jacobi polynomials, we show how our polynomials can be
 specialized to those cases. Henceforth, for notational
 convenience, we replace
$\tilde{\tilde{\,}}$ by
 $\tilde{\,}$ in the map (7.2).

\bigbreak

\example  {\bf $5$-step Laguerre polynomials}. Darboux
transforms for $L=\LR_{1}^{2}$, and weight $\rho_{0}
(z)=z^{\alpha} e^{-z} I_{[0,\iy)} (z)$ for $\al >0$.

Setting $m=2$ in formule (7.2), we find the map $$
\rho=(\rho_{0}(z), z\rho_{0}(z),
z^{2}\rho_{0}(z),\ldots) \mapsto \tilde{\rho} =
(\tilde{\rho_{0}}(z), \tilde{\rho_{1}}(z),
\tilde{\rho_{2}}(z), \ldots ), $$ with $$
\tilde{\rho_{0}}(z)=\Gamma(\alpha)(c\dt(z)+d
\dt'(z))+(b+\frac{e}{z}) \rho_{0}(z),~~\mbox{with}~b
\neq 0, $$ $$ \tilde{\rho_{i}}(z)=z^i\rho_{0}(z) =
z^{\alpha+i} e^{-z}
 I_{[0,\iy)}(z),\quad i\geq 1,
$$ obtained from formula (7.2), by setting, for
homogeneity considerations and without loss of
generality, $$ c_{0}=c\Gamma (\alpha),\quad
c_{1}=d~\Gamma(\alpha), \quad d_{0}=e, \quad d_{1}=b.
$$

The moments $\la z^i,\rho_j(z)\ra$ for the original
sequence are given by the following expressions $$
\mu_{ij}=\la z^i, \rho_j \ra=\la z^i, z^j \rho_0
\ra=\Gamma(\al+i+j+1)~,$$ with
polynomials\footnote{$\al!:=\Gamma(\al+1)$,
$(\al)_0=1$ and $(\al)_j=\al(\al-1)...(\al-j+1)$.}
\bean p_n^{(1)}(z)&=& \frac{1}{\det m_n}\pmatrix{\al!&
\left(\al+1\right)!&\left(\al+2\right)!&~~...~~&1\cr(\al+1)!
&\left(\al+2\right)!&\left(\al+3\right)!%
 &~~...~~&z\cr (\al+2)!&%
 \left(\al+3\right)!&\left(\al+4\right)!&~~...~~&z^{2}\cr %
 \left(\al+3\right)!&\left(\al+4\right)!&\left(\al+%
 5\right)!&~~...~~&z^{3}\cr \vdots%
 &\vdots&\vdots
 &~~...~~&\vdots\cr  \left(\al+n%
 \right)!&\left(\al+n+1\right)!&\left(\al+n+2\right)!
 &~~...~~&z^n\cr }\\
 &&\\
 &=& \sum_{i=0}^n \left(n\atop {i}\right)(\al+n)_i (-1)^i z^{n-i}
  ;\eean
 the latter are, as expected, the Laguerre polynomials orthogonal
 with regard to the weight $\rho_{0}(z)$.

The Darboux transformed moments $\tilde \mu_{ij}=\la
z^i,\tilde\rho_j(z)\ra$ are given by the following expressions
%$$
%\la z^0,\tilde \rho_0 \ra=(c+e)\Gamma(\al)+b\Gamma(\al+1)
%$$
\bean
\tilde \mu_{i0}&=&\la z^i,\tilde \rho_0 \ra=e\Gamma(\al+i)+b\Gamma(\al+i+1)
+(\dt_{i,0}~c -\dt_{i,1}~ d) \Gamma(\al),\\
\tilde \mu_{ij}&=&\la z^i,\tilde \rho_j \ra=
 \la z^i,z^j \rho_0 \ra=\Gamma(\al+i+j+1)~~~\mbox{for}~~~ j\geq 1
,\eean
from which one computes the Darboux transformed monic
polynomials

\bigbreak

\noindent$\displaystyle{(\det \tilde m_n)~\tilde p_n^{(1)}(z)}=$\hfill
\be\pmatrix{~~\left(\al-1\right)!\,e+
\al!\,b+\left(\al-1\right)!\,c&\left(\al+1%
 \right)!&\left(\al+2\right)!&~~...~~&1\cr ~~~\al!\,
 e+\left(\al+1\right)!\,b-\left(\al-1%
 \right)!\,d&\left(\al+2\right)!&\left(\al+3\right)!%
 &~~...~~&z\cr \left(\al+1\right)!\,e+\left(\al+2\right)!\,b&%
 \left(\al+3\right)!&\left(\al+4\right)!&~~...~~&z^{2}\cr %
 \left(\al+2\right)!\,e+\left(\al+3\right)!\,b&\left(\al+4\right)!&\left(a+%
 5\right)!&~~...~~&z^{3}\cr \vdots%
 &\vdots&\vdots
 &~~...~~&\vdots\cr  \left(\al+n-1\right)!\,e+\left(\al+n%
 \right)!\,b&\left(\al+n+1\right)!&\left(\al+n+2\right)!
 &~~...~~&z^n\cr }. \ee
 The appendix to this paper gives the first four $5$-step
 Laguerre polynomials.

The classical Laguerre polynomials are evidently
special cases of the Darboux transformed polynomials
$\tilde p^{(1)}_n$'s: $$p^{(1)}_n(z)=\tilde
p^{(1)}_n(z)|_{c=d=e=0,~b=1}. $$

It is interesting that, in an effort to find bispectral problems,
Gr\"unbaum and Haine \cite{GH} had obtained special cases of these
polynomials. Their method was to perform two explicit Darboux
transforms on the explicit square $L=\LR^2$
 of the
$3$-step relation  $\LR$ for the Laguerre polynomials. They found,
by computation, a new matrix $\tilde L$ and polynomials $\tilde
p(z)$, which coincide with ours,
 by
 setting $c=d=0, ~e/b=\al/r$ in (7.4), and hence $r\neq 0$. They show they
are related to Laguerre by means of a differential
equation. Indeed, given the differential equation for the
 Laguerre polynomials,
$$
B=-z \frac{\pl^2}{\pl z^2}+(z-\al-1)\frac{\pl}{\pl
z},~~\mbox{with}~~Bp_n(z)=np_n(z),
$$ and the operators
$$
P=B+\frac{\pl}{\pl z}+r~~\mbox{and}~~Q=B-\frac{\pl}{\pl z}+r+1 ,
$$ they show that the
 $ p^{(1)}_n$'s and $\tilde p^{(1)}_n$'s are
related by the following differential equations
$$
P p_n(z)=(n+r)\tilde p_n(z)~~\mbox{and}~~Q\tilde p_n(z)=(n+r+1)
p_n(z).
$$

\example {\bf $5$-step Jacobi polynomials}. ~~Darboux
transform for $L=\LR^2$ and Jacobi weight\footnote{It
is more convenient to base the Jacobi weight on [0,2]
rather then [-1,1].}
$\rho_0(z)=(2-z)^{\al}z^{\beta}I_{[0,2]}(z),$ for $\al
>-1$ and $\beta
>0$.
 Here the map is given by $\rho\longmapsto \tilde \rho$, with
\bea
 \tilde \rho_0(z)&=&\nu\left(c~\dt(z)+d
~\dt'(z)\right)+\rho_0(z)
\left(e+\frac{b}{z}\right),~~\mbox{with}~~e\neq
0\nonumber\\
  \tilde \rho_i(z)&=&z^i \rho_0(z)=
(2-z)^{\al}z^{\beta +i}
I_{[0,2]}(z)~~\mbox{for}~~i\geq 1, \eea with $$
\nu=2^{\al+\beta+1}\frac{\Gamma(\al+1)\Gamma(\beta
+1)}{\Gamma(\al+\beta+2)}.$$ As in the previous
example, the adjustments of constants was made for
homogeneity reasons.

The moments for the original sequence are given by
$$
\mu_{ij}=\la z^i, \tilde\rho_j\ra=2^{\al +\beta +i+j+1}
\frac{\al!(\beta +i+j)!}{(\al +\beta +i+j+1)!}~~
 \mbox{for}~~j\geq 1,
$$
and the Jacobi polynomials by

\bigbreak

\noindent$\displaystyle{p_n^{(1)}(z)= \frac{1}{\det m_n}}\times $\hfill
$$\det\pmatrix{{{\alpha!\,2^{\beta+\alpha+1}\,\beta!}\over{\left(\beta+%
 \alpha+1\right)!}}&{{\alpha!\,2^{\beta+\alpha+2}\,\left(\beta+1%
 \right)!}\over{\left(\beta+\alpha+2\right)!}}&{{\alpha!\,2^{\beta+%
 \alpha+3}\,\left(\beta+2\right)!}\over{\left(\beta+\alpha+3\right)!%
 }}&...&1\cr {{\alpha!\,2^{\beta+\alpha+2}\,\left(%
 \beta+1\right)!}\over{\left(\beta+\alpha+2\right)!}}&{{\alpha!\,2^{%
 \beta+\alpha+3}\,\left(\beta+2\right)!}\over{\left(\beta+\alpha+3%
 \right)!}}&{{\alpha!\,2^{\beta+\alpha+4}\,\left(\beta+3\right)!%
 }\over{\left(\beta+\alpha+4\right)!}}&...&z\cr {{%
 \alpha!\,2^{\beta+\alpha+3}\,\left(\beta+2\right)!}\over{\left(\beta%
 +\alpha+3\right)!}}&{{\alpha!\,2^{\beta+\alpha+4}\,\left(\beta+3%
 \right)!}\over{\left(\beta+\alpha+4\right)!}}&{{\alpha!\,2^{\beta+%
 \alpha+5}\,\left(\beta+4\right)!}\over{\left(\beta+\alpha+5\right)!%
 }}&...&z^{2}\cr \vdots&\vdots&\vdots&\vdots&\vdots%
 \cr {{\alpha!\,2^{\beta+\alpha+n+1}\,\left(\beta+n\right)!}\over{%
 \left(\beta+\alpha+n+1\right)!}}&{{\alpha!\,2^{\beta+\alpha+n+2}\,\left(%
 \beta+n+1\right)!}\over{\left(\beta+\alpha+n+2\right)!}}&{{\alpha!\,2^{%
 \beta+\alpha+n+3}\,\left(\beta+n+2\right)!}\over{\left(\beta+\alpha+n+3%
 \right)!}}&...&z^{n}\cr } $$
$$=\frac{1}{\det m_n}\sum^n_{k=0} (-2)^{n-k}\left( n \atop k  \right)
 (\al+\beta +n+k)_k (\beta +n)_{n-k} z^k .~~~~~~~~ $$
 The Darboux transformed moments are given by
$$
\la z^i, \tilde\rho_0\ra=2^{\al +\beta +i+1}
\frac{\al!(\beta +i)!}{(\al +\beta +i+1)!}
\left( (e+c\dt_{i0})+(b-d \dt_{i1})
\frac{\al +\beta +i+1}{2(\beta +i)}        \right)
$$
$$
\la z^i, \tilde\rho_j\ra=2^{\al +\beta +i+j+1}
\frac{\al!(\beta +i+j)!}{(\al +\beta +i+j+1)!}~
 ~\mbox{for}~~j\geq 1,
$$
and the new polynomials $\tilde p_n^{(1)}$ by:

\bigbreak

\noindent $\displaystyle{\tilde p_n^{(1)}=\frac{1}{\det m_n}\times\det }$\hfill

$$ \pmatrix{{{\alpha!\,2^{\beta+\alpha+1}\,\beta!\,\left(e+c+{{b\,%
 \left(\beta+\alpha+1\right)}\over{2\,\beta}}\right)}\over{\left(%
 \beta+\alpha+1\right)!}}&{{\alpha!\,2^{\beta+\alpha+2}\,\left(\beta+%
 1\right)!}\over{\left(\beta+\alpha+2\right)!}}&{{\alpha!\,2^{\beta+%
 \alpha+3}\,\left(\beta+2\right)!}\over{\left(\beta+\alpha+3\right)!%
 }}&...&1\cr {{\alpha!\,2^{\beta+\alpha+2}\,\left(%
 \beta+1\right)!\,\left(e+{{\left(\beta+\alpha+2\right)\,\left(b-d%
 \right)}\over{2\,\left(\beta+1\right)}}\right)}\over{\left(\beta+%
 \alpha+2\right)!}}&{{\alpha!\,2^{\beta+\alpha+3}\,\left(\beta+2%
 \right)!}\over{\left(\beta+\alpha+3\right)!}}&{{\alpha!\,2^{\beta+%
 \alpha+4}\,\left(\beta+3\right)!}\over{\left(\beta+\alpha+4\right)!%
 }}&...&z\cr {{\alpha!\,2^{\beta+\alpha+3}\,\left(%
 \beta+2\right)!\,\left(e+{{b\,\left(\beta+\alpha+3\right)}\over{2\,%
 \left(\beta+2\right)}}\right)}\over{\left(\beta+\alpha+3\right)!}}&%
 {{\alpha!\,2^{\beta+\alpha+4}\,\left(\beta+3\right)!}\over{\left(%
 \beta+\alpha+4\right)!}}&{{\alpha!\,2^{\beta+\alpha+5}\,\left(\beta+%
 4\right)!}\over{\left(\beta+\alpha+5\right)!}}&...&z^{2}
 \cr \vdots&\vdots&\vdots&\vdots&\vdots\cr
 {{\alpha!\,2^{\beta+\alpha+n+1}\,\left(\beta+n\right)!\,%
 \left(e+{{b\,\left(\beta+\alpha+n+1\right)}\over{2\,\left(\beta+n%
 \right)}}\right)}\over{\left(\beta+\alpha+n+1\right)!}}&{{\alpha!\,2^{%
 \beta+\alpha+n+2}\,\left(\beta+n+1\right)!}\over{\left(\beta+\alpha+n+2%
 \right)!}}&{{\alpha!\,2^{\beta+\alpha+n+3}\,\left(\beta+n+2\right)!%
 }\over{\left(\beta+\alpha+n+3\right)!}}&...&z^{n}}.%
 $$

Again in \cite{GH}, Gr\"unbaum and Haine had considered
 special cases of these polynomials. Namely,
the Jacobi polynomials satisfy a differential equation, $$
Bp_n^{(1)}=n(n+\al +\beta+1) p_n^{(1)} ,$$ involving the
differential operator
$$B=z(z-2)\left(\frac{\pl}{\pl z}\right)^2+\left( (\al +\beta +2)z-2(\beta +1)
\right)\frac{\pl}{\pl z}.
$$
Defining
$$
P=B-(z-2)\frac{\pl}{\pl z}+r~~\mbox{and}~~Q=B+(z-2) \frac{\pl}{\pl
z}+r+\al+\beta +1,
$$
they show the $p_n^{(1)}$ and $\tilde p_n^{(1)}$'s, for
$\displaystyle{c=0,d=0,e/b=r/2
 \beta}$ and hence $r\neq 0$, are related
by the following differential equations:
$$
Pp_n^{(1)}=(n^2+(\al +\beta)n+r)\tilde p_n^{(1)}
$$
$$
Q\tilde p_n^{(1)}=(n^2+(\al+\beta +2)n+\al +\beta +r+1)
 p_n^{(1)}.
$$

This paper shows that these polynomials have a determinantal
 representation in terms of moments, defined with respect to periodic
 sequences of weights.  Moreover, the vertex operator technology
 enables one to consider general $2m+1$-band matrices.
 It remains an interesting open question
 to investigate the differential equations satisfied by the
 general $2m+1$-step Laguerre or Jacobi polynomials.

\section{Appendix}

The first few $5$-step Laguerre polynomials are given by the
following polynomials, which, for convenience of notation,
 we did not make monic; set $\al=a$:
\begin{eqnarray*} \tilde p_1^{(1)}(z)&=&
\left(e+c+a\,b\right)\,z-a\,e+d-a^{2}\,b-a\,b \\ % &&\\
\tilde p_2^{(1)}(z)&=&
\left(2\,e+d+a\,c+2\,c+a\,b\right)\,z^{2}\\
&&-\left(4\,a\,e+6\,e+a^{2}%
 \,c+5\,a\,c+6\,c+2\,a^{2}\,b+4\,a\,b\right)\,z \\
 & &+\left(a+2\right)\,
 \left(2\,a\,e-a\,d-3\,d+a^{2}\,b+a\,b\right) \\
 && \\
 %\end{eqnarray*}
 %\begin{eqnarray*}
\tilde
p_3^{(1)}(z)&=&\left(6\,e+2\,a\,d+6\,d+a^{2}\,c+5\,a\,c+6\,c+2\,a\,b\right)\,z^{
3%
 }  \\
 &&-(18\,a\,e+48\,e+3\,a^{2}\,d+21\,a\,d+36\,d+2\,
 a^{3}\,c +18\,a%
 ^{2}\, c+\\
 &&\hspace{5cm}52\,a\,c+48\,c+6\,a^{2}\,b+18\,a\,b)\,z^{2}\\
 &&+\left(a+3%
 \right)\,\left(18\,a\,e+24\,e+a^{3}\,c+9\,a^{2}\,c+26\,a\,c
 +24\,c+6%
 \,a^{2}\,b+12\,a\,b\right)\,z\\
 &&-\left(a+2\right)\,\left(a+3\right)\,%
 \left(6\,a\,e-a^{2}\,d-7\,a\,d-12\,d+2\,a^{2}\,b+2\,a\,b
 \right) \\
&& \\ \tilde p_4^{(1)}(z)&=&
\left(24\,e+3\,a^{2}\,d+21\,a\,d+36\,d+a^{3}\,c+9\,a^{2}\,c+26\,a%
 \,c+24\,c+6\,a\,b\right)\,z^{4}\\
 &&-\bigl(96\,a\,e+360\,e+8\,a^{3}\,d+96%
 \,a^{2}\,d+376\,a\,d+480\,d+3\,a^{4}\,c \\
 &&~~~~~~~~~~~~~~~~+42\,a^{3}\,c+213\,a^{2}\,c+%
 462\,a\,c+360\,c+24\,a^{2}\,b+96\,a\,b\bigr)\,z^{3}\\
 &&+3\,\left(a+4%
 \right)\,\Bigl(48\,a\,e+120\,e+2\,a^{3}\,d+24\,a^{2}\,d+94\,a\,d+120%
 \,d+a^{4}\,c\\
 &&~~~~~~~~~~~~~~~~~+14\,a^{3}\,c+71\,a^{2}\,c+154\,a\,c+120\,c+12\,a^{2}\,b%
 +36\,a\,b\Bigr)\,z^{2}\\
 &&-\left(a+3\right)\,\left(a+4\right)\,\Bigl(96%
 \,a\,e+120\,e+a^{4}\,c+14\,a^{3}\,c+71\,a^{2}\,c+154\,a\,c
 \\
 &&\hspace{7cm}+120\,c+24%
 \,a^{2}\,b+48\,a\,b\Bigr)\,z\\
 && +\left(a+2\right)\left(a+3\right)%
 \left(a+4\right)\left(24ae-a^{3}d-12a^{2}d-47ad-60%
 d+6a^{2}b+6ab\right),
 \end{eqnarray*}
$
etc... $\hfill

\end{document}